\documentclass[11pt, draft, onecolumn ]{IEEEtran2}
\usepackage{color}     
\usepackage{epsf,psfrag,amssymb,amsfonts,latexsym,cite,verbatim,enumerate,subcaption}
\usepackage{srcltx,amsmath,cases,graphicx}
\usepackage{times, multirow, array}
\usepackage[mathscr]{eucal}

\def\psfancypar#1#2{\begingroup\def\par{\endgraf\endgroup\lineskiplimit=0pt}
               \setbox2=\hbox{\large\sc #2}
               \newdimen\tmpht \tmpht \ht2 \advance\tmpht by \baselineskip
               \font\hhuge=Times-Bold at \tmpht
               \setbox1=\hbox{{\hhuge #1}}
               \count7=\tmpht \count8=\ht1
               \divide\count8 by 1000 \divide\count7 by \count8 
               \tmpht=.001\tmpht\multiply\tmpht by \count7 
               \font\hhuge=Times-Bold at \tmpht
               \setbox1=\hbox{{\hhuge #1}}
               \noindent
                \hangindent1.05\wd1
               \hangafter=-2 {\hskip-\hangindent
               \lower1\ht1\hbox{\raise1.0\ht2\copy1}%
                \kern-0\wd1}\copy2\lineskiplimit=-1000pt}

\newcommand{\E}{\mbox{{\rm E}}}

\def\boxit#1{\vbox{\hrule\hbox{\vrule\kern3pt
        \vbox{\kern3pt#1\kern3pt}\kern3pt\vrule}\hrule}}

\def\reals{ { {\rm  I \kern-0.15em R }  } }
\def\complex{ {\,{{\rm C} \kern-0.50em \raise0.20ex {  |}}\, }}

\def\mubf{\hbox{\boldmath$\mu$\unboldmath}}

\def\Sigmabf{\hbox{$\bf \Sigma$}}

\def\Pibf{{\bf \Pi}}

\def\gbf{{\bf g}}
\def\hbf{{\bf h}}

\def\wbf{{\bf w}}
\def\xbf{{\bf x}}
\def\ybf{{\bf y}}

\def\xbf{{\bf x}}
\def\ybf{{\bf y}}
\def\Abf{{\bf A}}

\def\Hbf{{\bf H}}
\def\Ibf{{\bf I}}

\def\Rbf{{\bf R}}

\def\Wbf{{\bf W}}

\def\Cc{{\cal C}}

\def\Fc{{\cal F}}

\def\Kc{{\cal K}}
\def\Lc{{\cal L}}

\def\Nc{{\cal N}}

\def\Pc{{\cal P}}

\def\Rc{{\cal R}}
\def\Sc{{\cal S}}

\def\Wc{{\cal W}}

\def\be{\vskip .3cm \begin{equation}}
\def\ee{\end{equation} \vskip .4cm \noindent}
\newcommand{\R}{\mbox{$\hat {\bf R}_{N}$}}

\def\Rxx{\Rbf_{\ssstyle X\kern-.1em X}}

\let\ssstyle=\scriptscriptstyle

\def\Kout{\setbox1=\hbox{\Huge\bf K}\hbox to
1.05\wd1{\hspace{.05\wd1}
}}
\def\Sout{\setbox1=\hbox{\Huge\bf S}\hbox to 1.05\wd1{\hspace{.05\wd1}
}}

  \ifx\LabelFigloaded\MYundefined\relax
  \else
   \fi

  \def\LabelFigloaded{\relax}

  \chardef\LabelFigCatAt\the\catcode`\@
  \catcode`\@=11

 \let\LabelFigwlog@ld\wlog
 \def\wlog#1{\relax}

 \ifx\\\MYundefined@
    \let\\\relax
 \fi

  \def\ms@g{\immediate\write16}

 \def\N@wif{\csname newif\endcsname }
 \def\Temp@ {\N@wif\ifIN@}
 \ifx\INN@\MYundefined@
    \else \let\Temp@\relax
 \fi
 \Temp@

  \def\IN@{\expandafter\INN@\expandafter}
  \long\def\INN@0#1@#2@{\long\def\NI@##1#1##2##3\ENDNI@
    {\ifx\m@rker##2\IN@false\else\IN@true\fi}%
     \expandafter\NI@#2@@#1\m@rker\ENDNI@}
  \def\m@rker{\m@@rker}
 
  \newtoks\Initialtoks@  \newtoks\Terminaltoks@
  \def\SPLIT@{\expandafter\SPLITT@\expandafter}
  \def\SPLITT@0#1@#2@{\def\TTILPS@##1#1##2@{%
     \Initialtoks@{##1}\Terminaltoks@{##2}}\expandafter\TTILPS@#2@}

 \def\Shifted@@#1#2#3{\setbox0=\hbox{#3}%
   \raise -\dp0\vbox {\kern-#2%
       \hbox {\kern#1\unhbox0\kern-#1}%
           \kern#2}}

 \newcount\gridcount
 \newbox\auxGridbox@ \newbox\hGridbox@ \newbox\vGridbox@
 \newbox\Labelbox@ \newbox\auxLabelbox@
 \newbox\Coordinatebox@
 \newtoks\Labeltoks@
 \newdimen\Wdd@ \newdimen\Htt@
 \newdimen\Wddd@ \newdimen\Httt@
 
 \def\Wr@{\immediate\write16}

 \newdimen\GL@wd
 \def\GridLineWidth#1{\GL@wd=#1}

 \def\gobble#1{}
 \def\EdgeErr@{\Wr@{}%
      \Wr@{\string\Edges\space argument
      1, 10, 100 or 1000 please\string!}%
      }

 \newcount\Edgect@

 \def\Sweepup#1\endSweepup{}

 \def\SetEdges@{%
    \edef\Zr@@s{\expandafter\gobble\number\Edgect@\empty}%
        \count255=0\Zr@@s\relax
        \ifnum\count255=\z@\else\EdgeErr@\show\tailtest\fi
        \count255=1\Zr@@s\relax
        \ifnum\count255=\Edgect@\relax\else\EdgeErr@\show\leadtest\fi
    \EdgGl@b\edef\Zr@s{\expandafter\gobble\Zr@@s\empty}
    \ifnum\Edgect@>\@ne\relax\EdgGl@b\let\L@Dc\empty
        \else\EdgGl@b\edef\L@Dc{\string.}\fi
    \ifnum\Edgect@>\@ne\relax
        \EdgGl@b\edef\Edgescale@##1{\divide##1 by \Edgect@}%
        \else\EdgGl@b\edef\Edgescale@##1{}\fi
    }

 \def\Edges#1{\Edgect@=#1\relax
     \let\EdgGl@b\global \SetEdges@}

 \Edges{1}

 \def\hhrule{\hrule height \GL@wd\vskip-.\GL@wd}

 \def\hRule@{%
   \advance\gridcount -2%
   \vfil\hhrule\vfil
   \llap{\smash{\raise -2.5pt
     \hbox{\L@Dc\number\gridcount\Zr@s\kern2pt}}}%
   \hhrule
   }

\def\vvrule{\vrule width \GL@wd \kern-\GL@wd}

 \def\vRule@{\advance\gridcount 2%
   \hfil\vvrule\hfil
   \setbox\auxGridbox@=\vbox to 0pt
      {\vskip \Htt@\vskip 2pt
        \hbox to 0pt{\hss\L@Dc\number\gridcount\Zr@s\hss}\vss}%
      \wd\auxGridbox@=0pt \box\auxGridbox@
   \vvrule
   }

 \def\PlaceGrid@@{\gridcount=10 
  \setbox\hGridbox@=\hbox{%
        \hbox{%
             \hskip-.4pt\vrule
             \vbox to \Htt@{%
               \offinterlineskip\parindent=\z@\relax
               \hbox to \Wdd@{\hfil}
               \hRule@\hRule@\hRule@\hRule@
               \vfil\hhrule\vfil}%
             \vrule\hskip-.4pt}
    }%
  \gridcount=0%
  \setbox\vGridbox@=\hbox{%
      \vbox{\offinterlineskip\parindent=0pt\hsize=0pt
         \vskip-.4pt\hrule%
         \hbox to \Wdd@{%
                 \vtop to \Htt@{\vfil}%
                 \vRule@\vRule@\vRule@\vRule@
                 \hfil\vvrule\hfil}%
         \hrule\vskip-.4pt}}%
  \wd\hGridbox@=0pt\ht\hGridbox@=0pt
  \wd\vGridbox@=0pt\ht\vGridbox@=0pt
  \hbox{\box\hGridbox@\box\vGridbox@}%
  }

 \def\LabelsGlobal{\def\LabGl@b{\global}}
 \def\LabelsLocal{\def\LabGl@b{}}
 \LabelsGlobal 

 \def\SetLabels#1\endSetLabels{%
   \LabGl@b\Labeltoks@={#1()\\}%
   }

 \LabGl@b\Labeltoks@={()\\}

 \def\ShowGrid{\LabGl@b\let\PlaceGrid@\PlaceGrid@@}
 \def\HideGrid{\LabGl@b\let\PlaceGrid@\relax}
 \def\Grids{\ShowGrid\LabGl@b\let\GridSwitch@\ShowGrid}
 \def\noGrids{\HideGrid\LabGl@b\let\GridSwitch@\HideGrid}

 \noGrids

 \def\bAdjust@@{%
     \setbox\auxLabelbox@=\hbox{\raise \dp\auxLabelbox@
            \box\auxLabelbox@}}
 \def\bAdjust@{\let\vAdjust@\bAdjust@@}

 \def\eAdjust@@{\dimen0=-.5\ht\auxLabelbox@
     \advance\dimen0 by .5\dp\auxLabelbox@
     \setbox\auxLabelbox@=
            \hbox{\raise\dimen0\box\auxLabelbox@}}
 \def\eAdjust@{\let\vAdjust@\eAdjust@@}

 \def\tAdjust@@{%
     \setbox\auxLabelbox@=\hbox{\raise-\ht\auxLabelbox@
            \box\auxLabelbox@}}
 \def\tAdjust@{\let\vAdjust@\tAdjust@@}

 \let\vAdjust@\relax

 \def\lAdjust@{\let\hAdjust@\rlap}
 \def\rAdjust@{\let\hAdjust@\llap}

 \let\hAdjust@\relax\let\vAdjust@\relax

 \def\FetchLabel@#1(#2)#3\\{%
     \IN@0#2@@\ifIN@
        \setbox0=\hbox{\ignorespaces#1#3\unskip}%
        \ifdim\wd0>0pt
           \ms@g{}%
           \ms@g{ !!! Bad label(s)? !!!}%
           \message{ #1(#2)#3}%
        \fi
        \def\LabelMole@##1\endFetchLabel@{%
            \IN@0()\\@##1@%
            \ifIN@\def\Temp@{\FetchLabel@##1\endFetchLabel@}%
            \else\def\Temp@{}%
            \fi
            \Temp@
           }%
     \else
       \ignorespaces#1\unskip
       \setbox\auxLabelbox@=%
         \hbox to 0pt{\hss\ignorespaces\hAdjust@
          {\ignorespaces#3\unskip}\hss}%
       \vAdjust@
       \let\hAdjust@\relax\let\vAdjust@\relax
       \AugmentLabelBox@@{#2}%
       \ht\Labelbox@=0pt\dp\Labelbox@=0pt
       \let\LabelMole@\FetchLabel@%
     \fi\LabelMole@}

 \newtoks\XYSep@ 
 \def\SetXYSeparator#1{%
     \IN@0#1@@\ifIN@\XYSep@{*}%
     \else
     \XYSep@{#1}%
     \fi
     }

 \SetXYSeparator*

 \def\AugmentLabelBox@@#1{%
     \IN@0\the\XYSep@ @#1@\ifIN@
       \SPLIT@0\the\XYSep@ @#1@%
       \setbox\Labelbox@=\hbox to 0pt{%
         \unhbox\Labelbox@
         \Shifted@@{\the\Initialtoks@\Wddd@}%
         {\the\Terminaltoks@\Httt@}%
         {\box\auxLabelbox@}}%
     \else
         \ms@g{}%
         \ms@g{ !!! Bad insertion point. !!!}%
         \message{ (#1\ this point was rejected.)}%
     \fi
    }

 \def\FetchOption@#1[#2]#3\endFetchOption@{%
    \def\temp{#1}
    \ifx\temp\empty
       \Edgect@=#2\relax
       \let\EdgGl@b\relax
       \SetEdges@
       \Cleaner@#3%
    \fi}

 \def\Cleaner@#1[@]{\Labeltoks@{#1}}
     
 \def\PlaceLabels@@{\mathsurround=0pt
     \def\Cr@{\\}%
     \let\L\lAdjust@\let\R\rAdjust@
     \let\B\bAdjust@\let\E\eAdjust@\let\T\tAdjust@
     \expandafter\FetchOption@\the\Labeltoks@[@]\endFetchOption@
     \Wddd@=\Wdd@ \Edgescale@\Wddd@ 
     \Httt@=\Htt@ \Edgescale@\Httt@
     \expandafter\FetchLabel@\the\Labeltoks@\endFetchLabel@
     \box\Labelbox@
     }%

 \let \PlaceLabels@\PlaceLabels@@

 \def\AffixLabels#1{\setbox\Coordinatebox@=\hbox{#1}%
      \Wdd@=\wd\Coordinatebox@ \Htt@=\ht\Coordinatebox@
      \advance\Htt@ \dp\Coordinatebox@
      \hbox{\copy\Coordinatebox@\kern-\Wdd@ 
           \Shifted@@{0pt}{-\dp\Coordinatebox@}%
           {\PlaceLabels@\PlaceGrid@}%
           \kern\Wdd@}%
      \GridSwitch@ 
      \LabGl@b\Labeltoks@{()\\}%
      }
 
   \let\wlog\LabelFigwlog@ld   
   \catcode`\@=\LabelFigCatAt  

                                By

              Raymond S\'eroul <A18645@FRCCSC21.BITNET>
                                and 
              Laurent Siebenmann <lcs@topo.math.u-psud.fr>
    
              VERSIONS: July 1991, Oct 1991, Jan 1992, July 1992

INTRODUCTION

      This labelling package is intended for TeX users who
rely on non-TeX sources for for their graphics inserts.  It
provides means for adding TeX labels to such inserts with a
minimum of fuss. 

       For most labels, TeX users have in the past found it
reasonably convenient to rely on non-TeX sources. Typical
occasions when an inescapable need for TeX labels seemed to
arise are

 (a) when the graphics program lacks certain exotic or complex
mathematical symbols

 (b) when the very highest typographical quality is wanted for the
labels

 (c) when labels included with the graphics fail to print, 
 and you cannot figure out why (cf. boxedeps.doc).  The labels
 provided by labelfig.tex are 100

       Since this package first appeared, many users, who in the
past scarcely dreamed of using TeX labels, have come to use
nothing but.  So it is now appropriate to add

Intoxication Warning:  TeX labels may be addictive and expensive. 

     If you have a fast preview you may disagree, and even find
that this package provides an agreeable paste-up environment; see
extra applications at end.

     Note to publishers: It is possible and convenient to ultimately
export the TeX labels produced by labelfig.tex to become an integral
part of the EPS file. This is often desired by a publisher who typically
uses an "upmarket" graphics or page layout program, with which the
staff is skilled in perfecting figures.  See Appendix I for
a recipe.

     The authors are grateful to Patrick Ion of Math Reviews for
helpful comments and encouragement.

BASIC INSTRUCTIONS

    After reading in the macro file using

preview or proof your figure with a coordinate grid printed on
top, by typing the following:

    \ShowGrid  
    \AffixLabels{<the graphics insertion>}

Here <the graphics insertion> is what you would type to insert
the graphics object alone without the grid.  This must provide
for the space around it. For example <the graphics insertion>
might well be \BoxedEPSF{MyFigure scaled 700} using the
boxedeps.tex macro package (from same source); this provides a
TeX box containing the encapsulated PostScript insert specified by
the file MyFigure. \AffixLabels{...} provides the grid (supposing
\ShowGrid is present) and later, once you have specified labels
using the grid, it will "tack on" the labels.

     The grid is a sort of (usually elongated) checkerboard of
ten rows and ten columns and its (internal) partitions are by
default numbered  .1, ... ,.9  both horizontally (X-coordinate
running left to right) and vertically (Y-coordinate running bottom
to top).  Thus the points enclosed by the grid correspond to the
points of the unit square in the cartesian "X-Y" plane, the lower
left corner corresponding to the origin (0,0).  By extrapolation,
the full page corresponds to a larger rectangle in the plane.

     These coordinates serve to position labels as follows.
Before the \AffixLabels{...} command type label specifications:

  \SetLabels
   (<X-coordinate>*<Y-coordinate>) <first label> \\
   .
   .
   .
   (<X-coordinate>*<Y-coordinate>)  <last label> \\
  \endSetLabels

Each row specifies one label and is terminated by \\.  In each
row, the position indicator comes first; it is written as a
standard cartesian point except that the X- and Y- coordinates
are separated by * rather than a comma because TeX allows a
comma as decimal point. There are no dimension units to specify
as the unit is the grid itself.

     By default, this cartesian point specifies where the middle
of the baseline of the label will be located.  However if you precede
the point by \L [or \R] the left [or right] edge of the baseline will
be located there. Similarly you may also precede the point by \T, \E,
or \B to vertically align the top equator or bottom of the label box
at the specified point.  This gives nine standard positions of
the label with respect to the insertion point --- corresponding to
the eight principle points of the compas and the center

                     \L\T     \T      \R\T

                     \L\E     \E      \R\E

                     \L\B     \B      \R\B

But this neglects the default "baseline" level of TeX,
giving potentially three more positions

                     \L    <no tag>   \R

For text, the baseline level is often the preferred. Its relation to
the others is variable. It will often coincide with the bottom level,
as happens for "X".  But it is often distinct, as for "g", in which
case you have in all 12 distinct positions rather than 9.

     It is convenient to think of this specification of label
position as attaching the label by a thumb-tack to the coordinate
grid. There are up to twelve positions of the thumb-tack on the
label, while the position of the thumb-tack on the coordinate grid is
arbitrary.  Normally, one choses the position of the thumb-tack on
the label to be the one that is the closest to the item being
labeled.  There are good reasons for this "rule of thumb":

   (a)  It facilitates correct positioning at first try.

   (b)  If the scale of the figure must be altered after labels
have been affixed, the labels have a good chance of remaining well
positioned.

   (c)  The visible grid need not extend beyond the "bounding box"
for the figure, because the best preferred position is always
(at least almost) within the bounding box .

The second reason is particularly important. Indeed it often
happens that scale has to be altered after labelling begins, in
order to either provide space for the labels, or to adjust
proportions between the labels and the figure.  (The size of labels
is unaffected by scaling.)

     Here is an artificial but self-contained test which uses
TeX rules to make a graphics object.

TEST

    Do not skip this!

 \def\FrameIt#1{\hbox{\vrule$\vcenter {\hrule\kern3pt%
             \hbox {\kern3pt #1\kern3pt}%
               \kern3pt\hrule}$\relax\vrule}}

 \def\Caption#1#2{\FrameIt{%
       \vtop {\hsize=#1\relax \parindent=0pt
         \leftskip=0pt \rightskip=0pt plus15pt
         \parfillskip=0pt
         \lineskip=1pt\baselineskip=0pt
         #2}}}

 \def\FirstQuadrant{\hbox to 100pt{\vrule\vbox to 100pt{%
        \hbox to 100pt{\hfil}\vfil\hrule}\hss}}

  \SetLabels
    \R(.5*.2) $\zeta\,\cdot$\\
    (.9*-.10) $\xi$\\
    \R(-.03*.9) $\eta$\\
    \T(.5*.9) \Caption{70pt}{%
          \it The norm of
          $g(\xi+i\eta)$ is indicated on
          contours of this invisible surface.}\\
  \endSetLabels

  \AffixLabels{\FirstQuadrant}

  \end

  Note that the coordinates to use for labels are indicated on the
edges of the grid (when visible) corresponding to the conventional
x- and y- axes of the Cartesian plane. By default the grid is
1-by-1. However, by the command \Edges{100}, you can change this
to 100-by-100 and many users find this alternative most
convenient. Place the command \Edges{...} in your style file (or
header) since its effect is is global. Other possible edge values
are 10 and 1000.

  If you use the command \Edges{...} at all, do so with care.  For
if you accidentally delete an \Edges{...} command your labels will
abruptly be badly misplaced and may logically but mysteriously
generate "dimension too big" errors under TeX and "off page" errors
under your driver.  

  You can dictate the edgescale for an individual figure by giving
the scale in brackets immediately after \SetLabels.  Thus, to
import into an article using say \Edge{100} a figure labelled using
another edgescale, say the original 1-by-1 default, you can use
\SetLabels[1]...\endSetLabels.

GETTING IT DOWN PAT

     Complicated labeling deserves the same respect as
complicated mathematics.  Do not expect it to come out perfect the
first time!  What is needed in either case is a mechanism to
repeatedly typeset troublesome pieces.

     One mechanism is always available.  One does complicated
labelling in a separate "test" file involving just the figure being
labelled;  a texpert will know how to \dump TeX's current state as
a temporary format that restarts rapidly at each retry.  Usually,
one then pastes the completed labelled figure back into the main
TeX file, but, of course, one can also \input it as an auxiliary
file.

     If you do not have a TeXpert at handy, here is a first
approximation to an efficient setup. By deletions reduce a copy
of your article to just a few lines before and after the figure.
Now label the figure, and finally, copy and paste the labelled
figure to the original article. Then copy the next figure to label
into this testbed and repeat. The TeXpert can improve the  speed
at which TeX starts up, by compiling a format specifically for
your article; just one caution: best NOT include in the format
ephemeral details of setup like \Set<mydriver>ArtSpecials (from
boxedeps.tex because this reads  figure dimensions which you may
change during your work session.

     An improved mechanism to repeatedly typeset troublesome
pieces is now available on the Macintosh; it is called LinoTeX;
see the same ftp sources.  It could be set up on many types
of computer.

     Before using labelfig.tex to attach labels to a graphics
object inserted using boxedeps.tex or BoxedArt.tex, make it a
firm rule to carefully adjust the bounding box using the trimming
commands of these packages, and also at least tentatively scale
and position the object. Beware of changing the grid inadvertently
after the labels have been positioned.  For example, correcting
the bounding box of a PostScript graphics object can foul up the
labels by changing the coordinate grid to which the labels are
attached. This is particularly true for the trimming  commands of
boxedeps.tex and BoxedArt.tex. However, as noted already, change
of scale is much less disruptive, and modest adjustments should be
well tolerated.

     Sometimes the labels protrude so far from the bounding box
of a figure that the figure has to be repositioned.  Best do this
by ad hoc spacing, say using \hglue and \vglue; altering the
bounding box would create a vicious circle.

     Remember that you are responsible for preventing labels
from overlapping. You are responsible for all label typography
including size and style. A label is really just about anything
that can be put in a TeX box. Note that spaces at the beginning
and end of labels will normally be suppressed; if you really want
them you must protect them with TeX braces.

     This package temporarily sets the \mathsurround parameter
of TeX to zero  while the labels are being affixed. This is done
because nonzero \mathsurround space would influence the position
of left and right aligned labels; then, when a texpert or printer
modifies mathsurround, diagram labeling might be disastrously
altered. There is a small price to pay involving labels that are
formatted as caption boxes including mathematics: you  may want or
need to specify an explicit mathsurround space within the caption
box; it will not influence anything outside.

     Those hostile to the use of * as separator between
the X and Y coordinates of label insertion points, are free to
impose another using \SetXYSeparator{<the new separator>}.  
Americans may prefer "," to "*" since they never use a 
comma as a decimal point; on the other hand, * may be more visible.

APPENDIX (I)  MERGING labelfig.tex LABELS INTO AN EPSF GRAPHICS OBJECT.

     As promised in the introduction, here is a recipe useful for
publishers. It works at least on Macintosh and at least for vectorized
graphics and Adobe type1 fonts.  (There is surely a similar recipe for
PCs under MSWindows.)

 (a)  Use boxedeps.tex utility to integrate the figure given by the eps
file, "x.eps" say, with a visible frame around it.  See
\ShowDisplacementBoxes command in boxedeps.tex.  To get precise results
automatically it is important to use the \Trim... commands of
boxedeps.tex making the "DisplacementBox" neatly fit the figure.

 (b)  Use the TeX printer driver and LaserWriter (versions >= 8.1.1) to
export to an EPSF the DVI page containing the integrated, labelled
figure. You now have an EPS file  "xx.eps"  that contains too much, and at
the wrong scale, and at wrong position.

 (c)  Convert the EPSF to an Adode Illustrator format EPSF using
the shareware utility called epsConvert by Sam Weiss
1993-- (currently $25).

 (d)  In Illustrator (or a compatible program), group the labels and the
"DisplacementBox"; copy them to the clipboard and paste them into "x.ps".
This step requires that all the label fonts be "visible to the Macintosh.

 (e)  Translate and scale the pasted group consisting of the labels plus
the "DisplacementBox" so as to make the "DisplacementBox" the bounding
box of (labelless) figure represented by "x.eps".  At this point the
labels will be correctly placed on the figure "x.eps".

 (f)  Ungroup and delete the "DisplacementBox".  The result is the
desired single EPS file, "x+.eps" say, It contains the original figure
plus its labels.  

     Using grouping and ungrouping appropriately in "x+.eps", a
publisher's staff can very efficiently improve label positions etc.

APPENDIX II)  SOME EXOTIC APPLICATIONS

     The grid of labelfig.tex is analogous to a light-table in
classical page makeup with wax or latex glue.  In principle, you
can use it to compose any page from its indivisible parts.  This
even has some of the artisanal charm of classical paste-up
provided you have a fast screen preview to make the process
"interactive".

     In practice labelfig.tex is a tool for nonstandard jobs.
Here are a few going beyond the labelling already discussed.

(I)  GRAPHICS INTEGRATION.

     This is accomplished by treating the imported graphics
objects as labels.  The underlying graphics object is then
typically an empty  \vbox to <dimension>{\vfill} in a TeX
\midinsert...\endinsert construction.  A label line
might be of the form

   (.1*.1) \\

The exact form of the special command varies from driver to
driver.  However, in the case of encapsulated PostScript graphics
(EPSF norm), by relying on boxedeps.tex, one can have the
following standard syntax (independant of driver  (see
boxedeps.doc for details.
  
  (.1*.1) \BoxedEPSF{MyFigure scaled <scale in mils>}\\

This may be slow since it requires TeX to read the PostScript
file to read bounding box using many complex macros.  So you
may want to try

  (.1*.1) \EPSFSpecial{MyFigure}{<scale in mils>}\\

which is fast and driver independant, but it squashes the
bounding box, normally to its lower left corner.

     Similarly for graphics of the Macintosh PICT norm ---
using BoxedArt.tex (same sources) in place of boxedeps.tex.

     This approach to integration is to be recommended when
one is assembling a composite graphics object.

 (II)  COMMUTATIVE DIAGRAM ENHANCEMENT

     Commutative diagrams or arrays of mathematical objects
connected by arrows of various sorts are common in mathematics.
The mathematical objects require the use of TeX.  Recently TeX
acquired a good collection of arrows of all slopes --- that of
LamSTeX --- plus pwerful macros to build the diagrams.

     However, even the LamSTeX collection is often
inadequate; it lacks for example double shafted arrows, dotted
arrows and curved arrows. Fortunately it is possible to produce
such arrows on an individual basis using sophisticated graphics
programs such as Illustrator and AldusFreehand (both serving
the EPSF norm) or using Metafont (with its public domain norm).
Since the creation of each new arrow is a work of love, you
probably want to limit the number of arrows by using LamSTeX
for most arrows. The 40K commutative diagram module of LamSTeX
has been adapted to work with AmSTeX and a copy may be posted
with LabelFig and related files. Unfortunately no one has yet
offered a version that works with Plain TeX or LaTeX.

       Suffice it here to say that when the exotic arrow has
been somehow imported into TeX, labelfig.tex treats it as a
label that one affixes to the commutative diagram.  Two other
steps will be treated in separate notes, namely the matter of
extracting the dimension specifications for the arrow and the
construction of the arrow --- for these steps are far from
unique and often depend intimately on your computer environment. 
Notes for the Macintosh-Textures-Illustrator combination are
found in the file ExoticArrows.doc.

 (III) NESTING 

Ingenuity pays off in exploiting labelfig.tex. One can
mix graphics and typography quite freely.  labelfig.tex is good
for freeform or overlapping arrangements, while boxedeps.tex (or
BoxedArt.tex) is best for regimented non-overlapping
arrangements --- and the two can be combined.

     The default behavior of labelfig.tex is not ideal 
for nesting objects, because to prevent trouble for beginners
the register for labels is globally cleared when \AffixLabels
concludes.  But there are switches available

      \LabelsGlobal      \LabelsLocal

which change this.  To understand this, extend the above test 
by something like:

 \LabelsLocal

 \SetLabels
    (.5*.5) AAA\\
 \endSetLabels

 {
 \SetLabels
    (.5*.5) ZZZ\\
 \endSetLabels
   \AffixLabels{\FirstQuadrant}
 }

   \AffixLabels{\FirstQuadrant}

     There are however potential pitfalls.  Neither
labelfig.tex nor boxedeps.tex has been tested under extreme
conditions. Problems may occur if their procedures are
indiscriminately nested. For boxedeps.tex (not labelfig.tex)
there is a precise cause for worry, namely many of its
variables are "global", which means that TeX braces will not
provide the protection one might expect.

COMMAND SUMMARY FOR labelfig.tex

  Here [...] means optional (one or zero)
       [...]* means any number of such constructs

  \SetLabels
    [[<P>](<X><Sep><Y>) <label> \\]*
  \endSetLabels
  \ShowGrid  
  \AffixLabels{<the figure>}

   --- <P> is tack position, one of eleven or empty
              order irrelevant

                   \L\T      \T      \R\T

                   \L\E      \E      \R\E

                     \L               \R

                   \L\B      \B      \R\B

   --- (<X><Sep><Y>) insertion point;
  <Sep> is separator, = * by default;
  \SetXYSeparator{<Sep>} changes it.
   <X> and <Y> are real numbers

  --- <label> a label to attach 

  --- <the figure> the figure to label 

  \GlobalLabels (default)     
  \LocalLabels  setting for nested constructs.

 \Grids makes ALL grids appear; \HideGrid then makes just next disappear.
 \noGrids returns to default.  The commands are always global.

 \GridLineWidth{<dimension>} adjusts width of grid lines. Default is very
small, to give "hairline" effect. If your grid lines are missing try
setting \GridLineWidth{1pt}.

 \Edges#1 globally changes the edge size of all grids to the numerical 
value #1, which must be 1, 10, 100, or 1000.  The default is 1.

VERSION HISTORY.
 --- Jan 1993: \Edges#1 and [??] option after \SetLabels
 --- July 1992: \Grids, \noGrids, \HideGrid;
       Gridlines become hairlines; \GridLineWidth{<dimension>}.
 --- Oct 1991, Jan 1992: \SetXYSeparator{<Sep>},  \LabelsGlobal,
       \LabelsLocal.
 --- July 1991: first release

Address for bugs and other feedback:

        Raymond S\'eroul
        IREM and Lab. de Typographie Informatise
        Univ. Rene Descartes
        Strasbourg

    Tel 33-88-41-63-45
    Email:  A18645@FRCCSC21.BITNET

        Laurent Siebenmann
        Mathematique, Bat. 425,
        Univ de Paris-Sud,
        91405-Orsay,
        France

    Tel 33-1-6941-7949; 
    Email: lcs@topo.math.u-psud.fr

\def\scalefig#1{\epsfxsize #1\textwidth}

\def\tcr{\textcolor{red}}
\def\tcb{\textcolor{blue}}

\newtheorem{lemma}{Lemma}

\newtheorem{corollary}{Corollary}

\newtheorem{remark}{Remark}

\newtheorem{proposition}{Proposition}

\title{\huge{Beam Design and User Scheduling for Non-Orthogonal Multiple Access with Multiple Antennas Based on Pareto-Optimality}}

\author{
 Junyeong Seo, {\em Student~Member, IEEE}, Youngchul
Sung$^\dagger$\thanks{$^\dagger$Corresponding author}, {\em
Senior~Member, IEEE} \\
\thanks{The authors are with Dept. of Electrical Engineering,  KAIST, Daejeon 305-701, South
Korea. E-mail:jyseo@kaist.ac.kr, ysung@ee.kaist.ac.kr. }
}

\begin{document}

\maketitle

\begin{abstract}
In this paper, an efficient transmit beam design and user scheduling method is proposed for multi-user (MU) multiple-input single-output (MISO) non-orthogonal multiple access (NOMA) downlink, based on Pareto-optimality.  The proposed beam design and user scheduling method groups simultaneously-served users into multiple clusters with  practical two users in each cluster, and then applies spatical zero-forcing (ZF) across clusters to control  inter-cluster interference (ICI) and Pareto-optimal beam design with successive interference cancellation (SIC) to two users in each cluster to remove interference to strong users and leverage signal-to-interference-plus-noise ratios (SINRs) of interference-experiencing weak users. The proposed method has flexibility to control the rates of strong and weak users and numerical results show that the proposed method yields good performance.
\end{abstract}

\begin{keywords}
Non-orthogonal multiple access, multi-user MIMO, scheduling,
Pareto-optimal design, SIC
\end{keywords}

\section{Introduction}

NOMA is a promising technology for $5$G wireless networks to
increase the spectral efficiency\cite{Saito:13VTC}. Unlike
conventional orthogonal multiple access (OMA) which serves
multiple users based on time, frequency and/or spatial domains,
NOMA exploits the power domain that results from unequal
channel conditions under which users with strong
channels are basically limited by degree-of-freedom (DoF) such as bandwidth not by
noise but users with weak channels are limited by additive
noise\cite[P. 239]{Tse:book}. In NOMA with such channel conditions,
the base station (BS) uses superposition coding and allocates less
power to strong-channel users  and more power to weak-channel
users. Here, less power to strong users is not so detrimental since
strong users are in the DoF-limited regime, but more power to weak users
leverages received SINRs of weak users.  The strong
interference caused from more power assigned to weak users through good
channels to strong users is eliminated by SIC to maintain high
quality channels for strong users.

Initially,  NOMA with a single antenna in both the BS and users
was studied\cite{Saito:13VTC,Ding:14SPL,Timotheou:15SPL,
Liu15:PIMRC, Choi:14CL,Al:14ISWCS,So&Sung:15ComLet}. Recently,
there have been efforts to extend  NOMA to multiple-antenna
systems.
Unlike conventional MU multiple-antenna downlink systems which serve as many users as the number
of antennas, more users can be served in  multiple-antenna NOMA
systems. Although the possibility  that multiple-antenna NOMA can
outperform conventional multiple-antenna OMA was shown \cite{Lan14:ICSPCS,Chen:14ICCS},  many
important problems need to be investigated further for multiple-antenna NOMA. In
multiple-antenna NOMA systems, typically user grouping is
done first by forming clusters as many as the number of transmit
antennas and assigning multiple users to each cluster, and then multiple users in each cluster share the spatial dimension and are served in the power domain.
Since the performance of multiple-antenna NOMA significantly
depends on the channel conditions of users across clusters and
within each cluster, effective scheduling and user grouping methods
should be devised in order to achieve both MU diversity and the
NOMA gain from unequal channel conditions. In addition, the
problem of optimal beam design and power allocation compatible to
user scheduling should be solved to maximize the performance.

\subsection{Related Works}

There have been
several studies on multiple-antenna NOMA  for cellular downlink especially for MU-MISO downlink which is the main focus of this paper.
In \cite{Hanif16:TSP}, the downlink beam design for sum-rate maximization was considered for one given cluster based on
minorization-maximization. In \cite{Kim13:MILCOM,Liu15:ICCW,Sayed17:WD}, user scheduling and clustering is considered with the assumption that  two users are assigned to  each cluster.
In \cite{Kim13:MILCOM}, highly correlated users are chosen as candidates to
be clustered, and two users having the largest channel gain difference
are assigned to the same cluster. In \cite{Liu15:ICCW}, strong users
are selected by the semi-orthogonal user selection (SUS) algorithm \cite{Yoo&Goldsmith},
and then weak users are selected using the matching user selection algorithm
by considering inter-cluster interference. In \cite{Sayed17:WD},
a fairness-oriented user selection  algorithms was proposed
by selecting  two paired users based on their NOMA data rate. In all the works of \cite{Kim13:MILCOM,Liu15:ICCW,Sayed17:WD}, the beam design problem for MU-MISO NOMA was simplified by designing zero-forcing (ZF) beams based on strong users' channels and allocating the same beam to the weak user as the beam of the strong user in each cluster.
There also exists some study on multiple-input multiple-output (MIMO) NOMA. For example, in \cite{Ding:16TWC} the impact of user paring is analyzed with fixed power
allocation under the assumption that inter-cluster interference is removed by multiple receive antennas.
Some part of this work was included in \cite{Seo&Sung:17SPAWC}.

\subsection{Contributions}

In this paper, we
consider MU-MISO NOMA downlink with practical two users in
each cluster and solve the aforementioned problem for MU-MISO NOMA downlink. The
contributions of this paper are summarized in the below:

$\bullet$  First, we solved the Pareto-optimal beam design and power
allocation problem for two-user MISO broadcast channels (BCs) in
which superposition coding is used at the transmitter and the
interference at the strong user is eliminated by SIC while the
interference at the weak user is treated as noise. This  work is the
basis for successive development in this paper and is  valuable as
an independent item.

$\bullet$ We proposed an effective user scheduling, beam design and power allocation
method for MU-MISO NOMA downlink based on the above two-user
Pareto-optimal design result  by exploiting both the spatial domain provided by
multiple transmit antennas and the power domain provided by SIC.  The key advantages of the proposed method compared to the previous methods are that
1) the rates of strong users and weak users can be controlled arbitrarily under Pareto-optimality, which provides great operational flexibility to NOMA networks, and 2) beam design of the proposed method is generalized to include multi-dimensional subspace if available and to yield performance improvement, whereas the same one-dimensional beam is always used by both strong and weak users in the same cluster in the previous methods \cite{Kim13:MILCOM,Liu15:ICCW,Sayed17:WD}.

\textit{Notations:}  Vectors and matrices are written in boldface
with matrices in capitals. All vectors are column vectors. For a
matrix $\Abf$, $\Abf^*$, $\Abf^H$ and $\Abf^T$ indicate the
complex conjugate, conjugate transpose, and  transpose of $\Abf$,
respectively, and  $\Lc(\Abf)$ and $\Lc^\perp(\Abf)$ denotes the
linear space spanned by the columns of $\Abf$ and its orthogonal
complement, respectively.
 $\Pibf_\Abf$ and
$\Pibf_\Abf^\bot$ are the projection matrices to $\Lc(\Abf)$ and
$\Lc^\perp(\Abf)$,
 respectively. $||\xbf||$ represents the 2-norm of vector $\xbf$. $\Ibf$ denotes the identity matrix.
$\ybf\sim\Cc\Nc(\mubf,\Sigmabf)$ mean that random vector $\ybf$ is
circularly-symmetric complex Gaussian distributed with mean vector
$\mubf$ and covariance matrix $\Sigmabf$.

\section{System Model }
\label{sec:systemmodel}

We consider a single-cell MU-MISO NOMA downlink system with a BS
equipped with $N_t$ transmit antennas and $K$ single-antenna
users. We assume that the $K$ users in the cell are divided into
the set $\Kc_1$ of $K/2$ strong-channel users and the set $\Kc_2$
of $K/2$ weak-channel users.  We assume that out of the $K$ users
in the cell, $2K_c$ users are selected and simultaneously served
for each scheduling interval with $K_c \le N_t$ (thus $2N_t$ users can be served simultaneously) and  this
simultaneous service to $2K_c$ users is done by forming $K_c$
clusters with two paired users  in each cluster composed of one
from the strong user set $\Kc_1$ and the other from the weak user
set $\Kc_2$. We assume that the BS uses linear precoding and NOMA
is applied to two paired users in each cluster.  Under these
assumptions, the transmit signal of the BS for one scheduling
interval is given by
\begin{eqnarray}
    \xbf &=& \sum_{k=1}^{K_c} \left( \sqrt{p_1^{(k)}} \tilde{\wbf}_1^{(k)} s_1^{(k)} +  \sqrt{p_2^{(k)}} \tilde{\wbf}_2^{(k)} s_2^{(k)} \right),
\end{eqnarray}
where  $s_i^{(k)}$ is the transmit symbol for  user $i$ in cluster $k$
 from $\Cc\Nc(0,1)$, $\tilde{\wbf}_i^{(k)}$ is the $N_t \times 1$
beamforming vector for user $i$ in  cluster $k$  out of the
feasible beamforming vector set $\mathcal{W} := \{\tilde{\wbf} ~| ~
\|\tilde{\wbf}\|^2 \leq 1 \}$, and $p_i^{(k)}$ is the power assigned to
user $i$ in  cluster $k$. The total BS transmit power $P_T$
is equally divided into $P_T/K_c=P$ for each cluster.  Then, the
received signals of the two users in  cluster $k$ are
 given by
\begin{align}
     y_1^{(k)} &=  \tilde{\hbf}^{(k)H}_{1} \left( \sqrt{p_1^{(k)}}\tilde{\wbf}_1^{(k)} s_1^{(k)} + \sqrt{p_2^{(k)}}\tilde{\wbf}_2^{(k)} s_2^{(k)}\right) \label{eq:sysmodelrec1}\\
      &\hspace{-0.9em} + ~\tilde{\hbf}^{(k)H}_{1}
     \sum_{k' \ne k }^{K_c} \left( \sqrt{p_1^{(k')}} \tilde{\wbf}_1^{(k')} s_1^{(k')} + \sqrt{p_2^{(k')}} \tilde{\wbf}_2^{(k')} s_2^{(k')}  \right)
      + w_1^{(k)},  \nonumber  \\
     y_2^{(k)} &=    \tilde{\hbf}^{(k)H}_{2} \left( \sqrt{p_1^{(k)}}\tilde{\wbf}_1^{(k)} s_1^{(j)} + \sqrt{p_2^{(k)}}\tilde{\wbf}_2^{(k)} s_2^{(k)}\right)  \label{eqA:sysmodelrec2} \\
     & \hspace{-0.9em} + ~ \tilde{\hbf}^{(k)H}_{2}\sum_{k' \ne k }^{K_c} \left( \sqrt{p_1^{(k')}} \tilde{\wbf}_1^{(k')} s_1^{(k')} + \sqrt{p_2^{(k')}} \tilde{\wbf}_2^{(k')} s_2^{(k')}  \right) + w_2^{(k)}, \nonumber
\end{align}
where $\tilde{\hbf}_{i}^{(k)}$ denotes the {\em actual} $N_t \times 1$ (conjugated)
channel vector between the BS and user $i$ in cluster $k$, and
$w_i^{(k)}$ is the zero-mean additive white Gaussian noise (AWGN)
at user $i$ in  cluster $k$ from
$\mathcal{CN}(0,[\epsilon_{i}^{(k)}]^2)$.

In the MU-MISO-NOMA system, we have both the spatial
domain and the power domain.  We consider the design approach in
which  two users in each cluster are served in the power domain
with SIC while multiple clusters are served based on the spatial
domain.  Note that the last two terms in each of
the right-hand sides (RHSs) of \eqref{eq:sysmodelrec1} and
\eqref{eqA:sysmodelrec2} are the ICI and AWGN.
In order to control ICI, we apply spatial ZF
across clusters. However, due to lack of spatial dimensions,
we cannot remove ICI completely for all users. Hence, we remove
ICI for the strong users with spatial ZF
 to keep the strong users not interference-limited.
With this approach, the beam vector $\tilde{\wbf}_i^{(k)}$ can be expressed as
\begin{eqnarray}  \label{eq:systemmodelproj}
    \tilde{\wbf}_i^{(k)} = \Pibf^\bot_{\tilde{\Hbf}_k}\wbf_i^{(k)}, ~~i=1,2,
\end{eqnarray}
for some vector $\wbf_i^{(k)}$, where
\begin{equation}
\tilde{\Hbf}_k := [\tilde{\hbf}_1^{(1)}, \tilde{\hbf}_1^{(2)}, \cdots, \tilde{\hbf}_1^{(k-1)}, \tilde{\hbf}_1^{(k+1)}, \cdots, \tilde{\hbf}_1^{(K_c)}].
\end{equation}

Once the ICI is controlled and given, the model  \eqref{eq:sysmodelrec1} - \eqref{eqA:sysmodelrec2} is a two-user MISO BC.
Thus, our approach to the overall design is to first investigate the optimal beam
design and power allocation for a two-user MISO BC with SIC at the
strong user's receiver in Section \ref{subsec:MISO Broadcast channel with $2$ user},   to derive certain performance
properties relevant to selection of two users in a cluster in Section \ref{sec:perforamnce_analysis}, and
then to develop an overall user selection and beam design method for all
clusters with controlling ICI in Section \ref{sec:scheduling}.

\section{Two-user MISO broadcast channel with SIC: Pareto-optimal design}
\label{subsec:MISO Broadcast channel with $2$ user}

In this section,  we focus on optimal beam vector design and power
allocation for a two-user MISO BC with SIC at the strong user's receiver from the perspective of {\em
Pareto-optimality}.  With the cluster index $(k)$ omitted, the
two-user model  \eqref{eq:sysmodelrec1} -
\eqref{eqA:sysmodelrec2} for cluster $k$ is given by
\begin{equation}  \label{eq:effTwoUserModel}
   y_i =  \hbf^H_i (\sqrt{p_i}\wbf_i s_i +  \sqrt{p_j}\wbf_j s_j) + n_i  \quad i,j \in \{ 1,2\}, \;\;  j \neq i,
\end{equation}
where $p_1 + p_2 \leq P$ with  the total power $P$ allocated to
the cluster; $n_i \sim \mathcal{CN}(0,\sigma_i^2)$ is the sum of ICI and
AWGN;  and $\hbf_i$ is the {\em effective} channel for user $i$ (in cluster $k$) given by $\hbf_i = \Pibf^\bot_{\tilde{\Hbf}_k} \tilde{\hbf}_i^{(k)}$, $i=1,2$ from  \eqref{eq:sysmodelrec1},  \eqref{eqA:sysmodelrec2} and \eqref{eq:systemmodelproj} since
$\tilde{\hbf}_i^H \tilde{\wbf}_i = \tilde{\hbf}_i^H \Pibf^\bot_{\tilde{\Hbf}_k}\wbf_i = \tilde{\hbf}_i^H [\Pibf^\bot_{\tilde{\Hbf}_k}]^H\wbf_i = \hbf^H_i \wbf_i$.
The feasible set for $\wbf_i$ is given by $\Wc=\{\wbf ~|~ \|\wbf\|^2 \leq 1\}$ since the Pareto-optimal beam $\wbf_i$ under the model \eqref{eq:effTwoUserModel} lies in the linear space spanned by  $\hbf_1=\Pibf^\bot_{\tilde{\Hbf}_k} \tilde{\hbf}_1^{(k)}$ and $\hbf_2=\Pibf^\bot_{\tilde{\Hbf}_k} \tilde{\hbf}_2^{(k)}$\cite{Ho&Gesbert&Jorswieck:11arXiv},
 and hence $\|\wbf_i^{(k)}\|$ = $\|\Pibf^\bot_{\tilde{\Hbf}_k}\wbf_i^{(k)}\|=\|\tilde{\wbf}_i^{(k)}\|$ for Pareto-optimal beams.

Under the NOMA framework, we assume that user 1 is the
strong user and user 2 is the weak user, i.e., $\|\hbf_1\|^2/\sigma_1^2 >
\|\hbf_2\|^2/\sigma_2^2$ and that user 1 decodes the interference
from user 2 and subtracts it before decoding its own data while
user 2 treats the interference as noise. With this assumption, the
rates of the two users are given by
\begin{eqnarray}
     &&\hspace{-1.5em} R_1(\wbf_1, p_1)   = \log_2 \left(1 + \frac{s_1(\wbf_1, p_1)}{\sigma_1^2}\right)  \label{eq:data_rate1} \\
    &&\hspace{-1.5em}R_2(\wbf_1,\wbf_2, p_1, p_2) \nonumber \\
    &&\hspace{-1.5em}= \log_2 \left( 1 + \min\left\{ \frac{r_1(\wbf_2,p_2)}{s_1(\wbf_1, p_1) + \sigma_1^2} , \frac{s_2(\wbf_2, p_2)}{r_2(\wbf_1, p_1) + \sigma_2^2} \right\}\right), \nonumber
\end{eqnarray}
where the signal power and the interference power are respectively given by
\begin{equation}     \label{eq:signal_and_interference_power}
    s_i(\wbf_i, p_i) := p_i |\hbf_i^H\wbf_i|^2 \quad \text{and} \quad r_i(\wbf_j, p_j) := p_j |\hbf_i^H
    \wbf_j|^2.
\end{equation}
Note in \eqref{eq:data_rate1} that for
the rate of user 1, the interference from user 2 is not
incorporated due to SIC and the rate of user 2 is bounded by not
only the SINR of user 2 but also the required 'SINR' for user 1 to
decode the message of user 2 for SIC before decoding its own data.
Then, for the given (effective) channel vectors $(\hbf_1,\hbf_2)$, the
achievable rate region $\Rc$ of the two-user MISO-NOMA BC is
defined as the union of the rate-tuples that can be achieved by
all feasible beam  vectors and power allocation:
\begin{equation}
    \mathcal{R} := \bigcup\limits_{ \substack{(\wbf_1, \wbf_2) \in \mathcal{W}^2 \\p_1, p_2:~p_1,p_2\ge 0,\\ p_1+p_2=P}} ( R_1(\wbf_1, p_1),R_2(\wbf_1,\wbf_2, p_1, p_2 ) ).
\end{equation}

The \textit{Pareto boundary} of the rate region $\mathcal{R}$ is
the outer boundary of $\mathcal{R}$ for which the rate of any one
user cannot be increased without decreasing the rate of the other
user and  Pareto-optimality has been used widely as a general
optimal beam design criterion for MU-MISO networks with linear
precoding \cite{jorswieck:08WSA}.  A pair of beam vectors
$(\wbf_1,\wbf_2)$  not achieving a Pareto-boundary point is not
optimal since both users' rates can be increased by a better
designed beam pair. Note that the sum-rate optimal point is the
point on the Pareto-boundary where the Pareto-boundary and the minus
45$^o$ degree line touch in the $(R_1,R_2)$ plane, and the
Pareto-optimality provides a general optimality criterion because
we can change the rate operating point arbitrarily and optimally.   It is known
that the Pareto-boundary can be found by maximizing $R_2$ for each
given feasible $R_1^*$ \cite{jorswieck:08WSA}, i.e.,
\begin{eqnarray}
    \max\limits_{\substack{ \scriptstyle (\wbf_1, \wbf_2) \in \mathcal{W}^2 \\  \scriptstyle p_1, p_2: ~p_1,p_2 \ge 0, ~ p_1+p_2 = P}} && R_2(\wbf_1,\wbf_2, p_1, p_2) \nonumber\\
    \text{subject to}   && R_1(\wbf_1, p_i) = R_1^*. \label{eq:R1_constraint1}
\end{eqnarray}
By exploiting the relationship between the rates and the SINRs in
\eqref{eq:data_rate1}, the problem
 \eqref{eq:R1_constraint1} can be
rewritten in terms of SINR as
\begin{eqnarray}
    \max\limits_{\substack{ \scriptstyle (\wbf_1, \wbf_2) \in \mathcal{W}^2\\  \scriptstyle p_1, p_2: ~p_1,p_2\ge 0, \\p_1+p_2=P}} &&\gamma_2:=\min\left\{ \frac{r_1(\wbf_2,p_2)}{s_1(\wbf_1, p_1) + \sigma_1^2} , \frac{s_2(\wbf_2, p_2)}{r_2(\wbf_1, p_1) + \sigma_2^2} \right\} \nonumber\\
    \text{subject to}  ~~~ &&  \frac{s_1(\wbf_1, p_1)}{\sigma_1^2} = \gamma_1^*,  \label{eq:PO_constraint1_2}
\end{eqnarray}
where $\gamma_1^*$ is a given feasible target SINR for user $1$.
An efficient solution to the problem
\eqref{eq:PO_constraint1_2} exploits an efficient parameterization
of the beam vectors $\wbf_1$ and $\wbf_2$. Note that the number of
design variables in $\wbf_1$ and $\wbf_2$ is $2N_t$ complex
numbers. However, one can realize that it is sufficient that both
beam vectors are linear combinations of $\hbf_1$ and $\hbf_2$
(equivalently, $\Pi_{\hbf_2}\hbf_1$ and $\Pi^\bot_{\hbf_2}
\hbf_1$). A component in the beam vector not in the span of
$\hbf_1$ and $\hbf_2$ does not affect either the signal power or
the interference power and thus does not affect either $R_1$ or
$R_2$ \cite{Jorswieck}.  Thus, it is known from
\cite{Ho&Gesbert&Jorswieck:11arXiv} that  the Pareto-optimal beam
vectors for the  problem \eqref{eq:PO_constraint1_2} can be parameterized as
\cite{Ho&Gesbert&Jorswieck:11arXiv,Lindblom:13SP}
\begin{eqnarray}
    \wbf_1(\alpha_1, \beta_1) &=& \alpha_1 \frac{\Pi_{\hbf_2}\hbf_1}{\|\Pi_{\hbf_2}\hbf_1\|} + \beta_1 \frac{\Pi^\bot_{\hbf_2} \hbf_1}{\|\Pi^\bot_{\hbf_2}\hbf_1\|}, \label{eq:parameterization1}\\
    \wbf_2(\alpha_2) &=& \alpha_2  \frac{\Pi_{\hbf_1}\hbf_2}{\|\Pi_{\hbf_1}\hbf_2\|} + \sqrt{1 - \alpha_2^2}  \frac{\Pi^\bot_{\hbf_1} \hbf_2}{\|\Pi^\bot_{\hbf_1}\hbf_2\|} \label{eq:parameterization2} ,
\end{eqnarray}
where $ (\alpha_1, \beta_1) \in \Fc  := \{(\alpha,\beta),
\alpha,\beta \geq 0, \alpha^2 + \beta^2 \leq 1 \} ~~\mbox{and}~~
\alpha_2 \in [0,1]$.
 Unlike the
conventional parametrization without SIC in which both users use
full power\cite{jorswieck:08WSA,Lindblom&Larsson11ICASSP}, in the
parameterization \eqref{eq:parameterization1} -
\eqref{eq:parameterization2}  user 1 may not use full power
whereas user 2 uses full power. This is because full power use of
user 2 helps both SIC at user 1 and its own SINR at user 2, but
full power use of user 1 is beneficial for its own rate but
detrimental to user 2's rate since user 2 treats interference as
noise. Substituting \eqref{eq:parameterization1} -
\eqref{eq:parameterization2} into
\eqref{eq:signal_and_interference_power}, we have
\begin{eqnarray}
    s_1(\wbf_1) &=& p_1\left( \alpha_1 \|\Pi_{\hbf_2}\hbf_1\| + \beta_1 \|\Pi^\bot_{\hbf_2}\hbf_1\|\right)^2 \nonumber\\
                &=& p_1 \|\hbf_1\|^2( \sqrt{\theta} \alpha_1 +  \sqrt{1- \theta} \beta_1)^2 \label{eq:s_1_parameterization}\\
    r_2(\wbf_1) &=&  p_1 \alpha_1^2 \frac{|\hbf_2^H \hbf_1|^2}{\|\Pi_{\hbf_2}\hbf_1\|^2} = p_1 \|\hbf_2\|^2 \alpha_1^2 \label{eq:r_2_parameterization} \\
    s_2(\wbf_2) &=& p_2 \|\hbf_2\|^2(\sqrt{\theta} \alpha_2 + \sqrt{1 - \theta} \sqrt{1 - \alpha_2^2})^2 \label{eq:s_2_parameterization} \\
    r_1(\wbf_2) &=&  p_2 \alpha_2^2 \frac{|\hbf_2^H \hbf_1|^2}{\|\Pi_{\hbf_1}\hbf_2\|^2} =  p_2 \|\hbf_1\|^2 \alpha_2^2, \label{eq:r_1_parameterization}
\end{eqnarray}
where the angle parameter $\theta$ between two effective channel vectors  $\hbf_1$ and $\hbf_2$ is defined as
\begin{equation}  \label{eq:angletheta}
\theta := \frac{|\hbf_1^H \hbf_2|^2}{\|\hbf_1\|^2 \|\hbf_2\|^2} ~~\in~ [0,~1].
\end{equation}
Substituting  \eqref{eq:s_1_parameterization} - \eqref{eq:r_1_parameterization} into the problem \eqref{eq:PO_constraint1_2} and taking square-root operation yield

\begin{eqnarray}
    \max\limits_{\substack{ \scriptstyle (\alpha_1,\beta_1) \in \Fc\\  \alpha_2 \in [0,1]\\ \scriptstyle 0 \leq p_1 \leq P}}
    && \gamma_2= \min\left\{ \frac{ \sqrt{P-p_1} \|\hbf_1\| \alpha_2}{\sqrt{\sigma_1^2(1+\gamma_1^*)}} ,
    \frac{\sqrt{P-p_1} \|\hbf_2\|(\sqrt{\theta} \alpha_2 + \sqrt{1 - \theta} \sqrt{1 - \alpha_2^2})}{\sqrt{p_1 \|\hbf_2\|^2 \alpha_1^2 + \sigma_2^2}} \right\} \label{eq:PO_objective_3}\\
    \text{subject to}   && \sqrt{p_1} \|\hbf_1\|( \sqrt{\theta} \alpha_1 +  \sqrt{1- \theta} \beta_1)  = \sqrt{\gamma_1^* \sigma_1^2}. \label{eq:PO_constraint1_3}
\end{eqnarray}

For later use, we define the following channel quality factor $\lambda_i$ and the normalized target SINR value for user 1, $\Gamma$:
\begin{equation} \label{eq:channelSNRandGamma}
\lambda_{i}:= \frac{||\hbf_i||^2}{\sigma_i^2}, ~i=1,2,
~~\mbox{and}~~ \Gamma := \gamma_1^*/\lambda_1.
\end{equation}
Here, $\lambda_i$ indicates the signal-to-noise ratio (SNR)
quality of user $i$'s channel, whereas $\theta$ in
\eqref{eq:angletheta} is a measure of the angle between the two
users' channels. Note that the actual target SINR for user 1
$\gamma_1^*$ is given by $\gamma_1^* = \Gamma \lambda_1$ and the
feasible range for $\Gamma$ is $\Gamma \in [0, P]$, where the
maximum $P$ occurs when $\wbf_1=\hbf_1/||\hbf_1||$ and $p_1=P$
since $\gamma_1 = s_1(\wbf_1,p_1)/\sigma_1^2 =
p_1|\hbf_1^H\wbf_1|^2/\sigma_1^2$.

The optimization problem \eqref{eq:PO_objective_3} -
\eqref{eq:PO_constraint1_3} with fixed power allocation
$p_1=p_2=1$ was solved in  \cite{Lindblom:13SP} under the
framework of a two-user  MISO interference channel. In the MISO
interference channel case, two transmitters in the network neither
cooperate nor share transmit power and hence the two transmit
power values $p_1$ and $p_2$ are fixed. On the other hand, in the
MISO-BC case the two power values $p_1$ and $p_2$ can be designed
at the BS under the constraints  $p_1,p_2 \ge 0$ and $p_1+p_2 = P$
to maximize the performance. Since our solution to the two-user
MISO-NOMA-BC case is based on the result from
\cite{Lindblom:13SP}, we briefly introduce the relevant result in
\cite{Lindblom:13SP} for further development in later sections.

\subsection{Background: The fixed power allocation case \cite{Lindblom:13SP}}
\label{sec:fixed_power_allocation}

In this subsection, we fix $p_1=p_2=1$ and  follow
\cite{Lindblom:13SP}. First, note that $\alpha_1$ appears only in
the constraint \eqref{eq:PO_constraint1_3} and  the denominator in
the second term in the RHS of \eqref{eq:PO_objective_3}. Thus,
optimal $\alpha_1$ can be found by solving the following
problem\cite{Lindblom:13SP}:
\begin{eqnarray}
    \min\limits_{\substack{ \scriptstyle (\alpha_1,\beta_1) \in \Fc}} && \alpha_1 \label{eq:PO_objective_41}\\
    \text{subject to}   &&  \|\hbf_1\|( \sqrt{\theta} \alpha_1 +  \sqrt{1- \theta} \beta_1)  = \sqrt{\gamma_1^* \sigma_1^2}, \label{eq:PO_constraint1_42}
\end{eqnarray}
This is because  the  second term in the RHS of
\eqref{eq:PO_objective_3} decreases monotonically with respect to
$\alpha_1$ and  hence maximizing the  second term in the RHS of
\eqref{eq:PO_objective_3} is equivalent to minimizing $\alpha_1$.
The problem \eqref{eq:PO_objective_41} -
\eqref{eq:PO_constraint1_42} can  easily be solved based on the
relationship  between a line segment \eqref{eq:PO_constraint1_42}
and the unit-radius ball $\Fc$ and the solution is given by
\cite{Lindblom:13SP}
\begin{equation}    \label{eq:x_1_opt}
    \alpha_1^* = \left\{ \begin{array}{cc}
                      0 & \quad   \text{if} \quad \Gamma \le 1 - \theta\\
                      \sqrt{\theta \Gamma} - \sqrt{(1-\theta) (1 - \Gamma )} & \quad  \text{if} \quad  \Gamma > 1 -
                      \theta,
                    \end{array}
    \right.
\end{equation}
where $\Gamma$ is defined in \eqref{eq:channelSNRandGamma} and
$\Gamma \in [0,1]$ for $p_1=1$. With the optimal $\alpha_1^*$ in
\eqref{eq:x_1_opt}, the corresponding optimal $\beta_1^*$ can be
found by the constraint \eqref{eq:PO_constraint1_3}, and
substituting  the optimal $\alpha_1^*$ into the problem
\eqref{eq:PO_objective_3} - \eqref{eq:PO_constraint1_3} yields
\cite{Lindblom:13SP}
\begin{equation}  \label{eq:PO_objective_6}
    \max_{\alpha_2 \in [0,1]}  \gamma_2=\min\left\{  a \alpha_2, ~b \alpha_2 + c \sqrt{1 -\alpha_2^2}\right\}
\end{equation}
where
$a :=\frac{\|\hbf_1\|}{\sqrt{\sigma_1^2 (1+ \gamma_1^*)}}$, $b :=
\frac{\|\hbf_2\| \sqrt{\theta}}{\sqrt{\|\hbf_2\|^2
(\alpha_1^*)^{2} + \sigma_2^2}}$,  and  $c := \frac{\|\hbf_2\| \sqrt{1- \theta}}{\sqrt{\|\hbf_2\|^2
(\alpha_1^*)^{2} + \sigma_2^2}}$.

\begin{figure}[htbp]
\centerline{
 \begin{psfrags}
 \psfrag{c1}[c]{\small case 1} %
  \psfrag{c2}[c]{\small case 2} %
  \psfrag{c3}[c]{\small case 3} %
  \psfrag{1}[c]{\small $1$} %
  \psfrag{a2}[l]{\small $\alpha_2$} %
  \psfrag{ap}[c]{\small $\alpha_2'$} %
  \psfrag{aa1}[l]{\small \tcb{$a\alpha_2$}} %
  \psfrag{aa2}[l]{\small \tcb{$a\alpha_2$}} %
  \psfrag{aa3}[l]{\small \tcb{$a\alpha_2$}} %
  \psfrag{b}[c]{\small \tcr{$b\alpha_2+c\sqrt{1-\alpha_2^2}$}} %
    \scalefig{0.45}\epsfbox{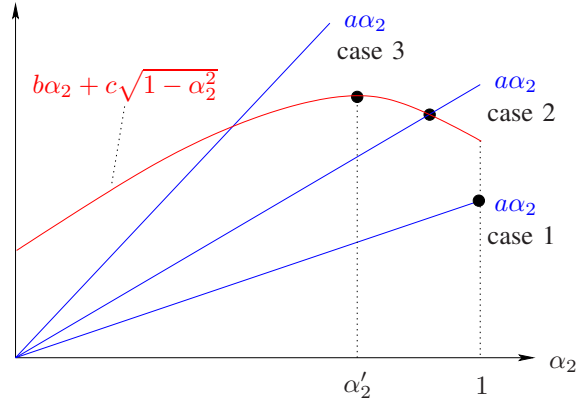}
\end{psfrags}
} \caption{ $a \alpha_2$ and $b\alpha_2 + c\sqrt{1-\alpha_2^2}$
(figure adapted from Fig. 1 in \cite{Lindblom:13SP})}
\label{fig:2userMISOICSICgraph}
\end{figure}

Note that the graph $(\alpha_2,a\alpha_2)$ of the first term
$a\alpha_2$ in the minimum in \eqref{eq:PO_objective_6} is a
straight line and the graph $(\alpha_2,b\alpha_2+c\sqrt{1 -
\alpha_2^2})$ of the second term $b\alpha_2+c\sqrt{1 -\alpha_2^2}$
in the minimum in \eqref{eq:PO_objective_6} is a straight line
plus a quarter circle, as shown in Fig.
\ref{fig:2userMISOICSICgraph}. Note also that $b\alpha_2 +
c\sqrt{1-\alpha_2^2}$ is maximized at $\alpha_2'=b/\sqrt{b^2+c^2}$
with maximum $\sqrt{b^2+c^2}$, and the intersection of $a\alpha_2$
and $b\alpha_2+c\sqrt{1-\alpha_2^2}$ occurs at
$\alpha_2''=c/\sqrt{c^2+(a-b)^2}$. There exist three different
cases for the solution to \eqref{eq:PO_objective_6} depending on
the relationship between the two graphs, as shown in Fig.
\ref{fig:2userMISOICSICgraph}. In the first case of $a \le b$,
$a\alpha_2$ is below $b\alpha_2+c\sqrt{1 -\alpha_2^2}$, the
minimum of the two is $a \alpha_2$, and thus optimal $\alpha_2^*$
is $1$. In the second case that $a\alpha_2$ and
$b\alpha_2+c\sqrt{1 -\alpha_2^2}$ intersect between $\alpha_2'$
and $1$, i.e., $\alpha_2' \le \alpha_2''$, the solution to
\eqref{eq:PO_objective_6} occurs at $\alpha_2^*=
\alpha_2''=c/\sqrt{c^2+(a-b)^2}$. Finally, in the third case that
$a\alpha_2$ and $b\alpha_2+c\sqrt{1 -\alpha_2^2}$ intersect
between $0$ and $\alpha_2'$, i.e., $\alpha_2' > \alpha_2''$, the
solution to \eqref{eq:PO_objective_6} occurs at $\alpha_2^*=
\alpha_2'=b/\sqrt{b^2+c^2}$. Since the condition $\alpha_2' \le
\alpha_2''$ can be rewritten as \cite{Lindblom:13SP}
\begin{equation}
       \frac{b}{\sqrt{b^2 + c^2}} \le \frac{c}{\sqrt{c^2 + (a-b)^2}}   \Longleftrightarrow a \leq b
     +c^2/b,
\end{equation}
 the optimal solution $\alpha_2^*$
to the problem \eqref{eq:PO_objective_6} is  summarized as
\cite{Lindblom:13SP}
\begin{equation}
    \alpha_2^*  = \left\{ \begin{array}{ll}
                       1 & \quad \text{if } a \le b~(\mbox{case 1}),\\
                       \frac{c}{\sqrt{c^2 + (a-b)^2}} & \quad  \text{if } b < a \leq b + c^2/b~(\mbox{case 2}), \\
                       \frac{b}{\sqrt{b^2+c^2}}=\sqrt{\theta}  & \quad \text{if } a > b + c^2/b~(\mbox{case 3}),
                     \end{array}
     \right.
\end{equation}
and the corresponding optimal value $\gamma_2^*$ is given by \cite{Lindblom:13SP}
\begin{equation}
    \gamma_2^* = \left\{ \begin{array}{ll}
         \gamma_2^{*(1)}=  a^2=\frac{  \|\hbf_1\|^2 }{\sigma_1^2(1+\gamma_1^*)},
                              & \quad \text{case 1}, \\
         \gamma_2^{*(2)}=       a^2(\alpha_2^*)^2=\frac{  \|\hbf_1\|^2 }{\sigma_1^2(1+\gamma_1^*)} (\alpha_2^*)^{2},
                              & \quad \text{case 2}, \\
         \gamma_2^{*(3)}=   b^2+c^2=\frac{ \|\hbf_2\|^2}{\|\hbf_2\|^2 (\alpha_1^*)^{2} + \sigma_2^2}, & \quad \text{case 3}.
                         \end{array}
       \right. \label{eq:gamma_2_fixed_power}
\end{equation}

\subsection{Pareto-optimal design in two-user MISO BC with SIC with power allocation}
\label{sec:PA_problem}

Now consider the actual problem \eqref{eq:PO_objective_3} -
\eqref{eq:PO_constraint1_3} of Pareto-optimal beam design and
power allocation for the two-user MISO-NOMA BC. We obtain the
solution to this problem  by  exploiting the result in Section
\ref{sec:fixed_power_allocation}. Note that once the power
allocation values  $p_1$ and $p_2$ are fixed, the corresponding
optimal solution can be obtained from the result in Section
\ref{sec:fixed_power_allocation}.  Therefore, we represent the
optimal solution as a function of power allocation  and then
optimize the power allocation. For given $p_1 \in [0,P]$, the
problem \eqref{eq:PO_objective_41} - \eqref{eq:PO_constraint1_42}
changes to
\begin{eqnarray}
    \min\limits_{\substack{ \scriptstyle (\alpha_1,\beta_1) \in \Fc}} && \alpha_1 \label{eq:PO_objective_4}\\
    \text{subject to}   &&  \sqrt{p_1}\|\hbf_1\|( \sqrt{\theta} \alpha_1 +  \sqrt{1- \theta} \beta_1)  = \sqrt{\gamma_1^* \sigma_1^2}, \nonumber
\end{eqnarray}
and the corresponding solution is given by   $\alpha_1^*(p_1) =$
\begin{equation}
    \left\{ \begin{array}{cc}
                      0 & \quad   \text{if} \quad \Gamma  \le   p_1(1 - \theta),\\
                      \sqrt{\theta \Gamma/p_1} - \sqrt{(1-\theta) (1 - \Gamma/p_1 )} & \quad  \text{if} \quad  \Gamma > p_1(1 -
                      \theta).
                    \end{array}
    \right. \label{eq:x_1_opt_with_p1}
\end{equation}
Furthermore, the coefficients $a$, $b$, and $c$ defined in \eqref{eq:PO_objective_6}
 are changed to
\begin{eqnarray}
    a(p_1) &:=&   \sqrt{P - p_1} \frac{ \|\hbf_1\|}{\sqrt{\sigma_1^2 (1+ \gamma_1^*)}},  \label{eq:a_p1}\\
    b(p_1) &:=&   \sqrt{P - p_1} \frac{\|\hbf_2\| \sqrt{\theta}}{\sqrt{p_1\|\hbf_2\|^2  (\alpha_1^*(p_1))^{2} + \sigma_2^2}}, \label{eq:b_p1} \\
    c(p_1) &:=&   \sqrt{P - p_1} \frac{  \|\hbf_2\| \sqrt{1- \theta}}{\sqrt{p_1\|\hbf_2\|^2 (\alpha_1^*(p_1))^{2} + \sigma_2^2}}. \label{eq:c_p1}
\end{eqnarray}
Then,  the optimal solution to
\eqref{eq:PO_objective_3} - \eqref{eq:PO_constraint1_3} can be
represented as a function of $p_1$:
\begin{equation}
    \alpha_2^*(p_1)  = \left\{ \begin{array}{ll}
                       1 & \quad \text{if } p_1 \in \mathcal{P}_1,\\
                       \frac{c(p_1)}{\sqrt{c^2(p_1) + [a(p_1)-b(p_1)]^2}} & \quad  \text{if } p_1 \in \mathcal{P}_2, \\
                       \frac{b(p_1)}{\sqrt{b^2(p_1)+c^2(p_1)}} =\sqrt{\theta} & \quad \text{if } p_1 \in
                       \mathcal{P}_3,
                     \end{array}
     \right. \label{eq:x_2_optimal_with_p1}
\end{equation}
where  $\mathcal{P}_1 := \{ p_1|  a(p_1) \le b(p_1) \}$,
$\mathcal{P}_2 := \{ p_1|  b(p_1) < a(p_1) \leq b(p_1) +
c^2(p_1)/b(p_1) \}$, and $\mathcal{P}_3 := \{ p_1|  a(p_1) >
b(p_1) + c^2(p_1)/b(p_1) \}$, and the corresponding SINR for user
2 $\gamma_2^{*}(p_1)$ is given by $ \gamma_2^*(p_1) = $
\begin{equation} \label{eq:gamma2_p1}
   \left\{ \begin{array}{ll}
                           \gamma_2^{*(1)}(p_1) = (P-p_1)  \frac{ \|\hbf_1\|^2 }{\sigma_1^2(1+\gamma_1^*)}   & \quad \text{if }  p_1 \in \mathcal{P}_1, \\
                           \gamma_2^{*(2)}(p_1) = (P-p_1) \frac{  \|\hbf_1\|^2 }{\sigma_1^2(1+\gamma_1^*)} [\alpha_2^{*}(p_1)]^2 & \quad \text{if } p_1 \in \mathcal{P}_2, \\
                           \gamma_2^{*(3)}(p_1) = (P-p_1) \frac{  \|\hbf_2\|^2}{\|\hbf_2\|^2 p_1 [\alpha_1^{*}(p_1)]^2 + \sigma_2^2} & \quad \text{if }  p_1 \in
                            \mathcal{P}_3.
                         \end{array}
       \right.
\end{equation}
Finally,  the original problem \eqref{eq:PO_objective_3} -
\eqref{eq:PO_constraint1_3} reduces to
\begin{equation} \label{eq:simplifed_problem}
    \max_{0 \leq p_1 \leq P} \gamma_2^*(p_1),
\end{equation}
where $\gamma_2^*(p_1)$ is given by  \eqref{eq:gamma2_p1}.
Note
that if we knew which $\gamma_2^{*(i)}$ in \eqref{eq:gamma2_p1} to use for optimization by directly
computing $a(p_1)$, $b(p_1)$, $c(p_1)$ and their relationship, it
would be easy to solve the problem \eqref{eq:simplifed_problem}. However, the  parameters $a(p_1)$, $b(p_1)$ and $c(p_1)$
 to determine $\gamma_2^{*(i)}$ to use  are functions of the design variable $p_1$. Nevertheless, this is possible and the result is given in the following proposition.

\vspace{0.5em}

\begin{proposition} \label{pro:PcregionDetermine}
For given $\hbf_1$, $\hbf_2$, $\sigma_1^2$, $\sigma_2^2$, $P$ and  $\gamma_1^*$, we have the
following regarding  the optimal solution $p_1^{opt}$ to the problem
\eqref{eq:simplifed_problem}. If $\theta \Gamma  < \tau$ or if  $\theta \Gamma \geq \tau \geq 0$ and $P \geq \Gamma + \frac{1}{1- \theta} (\sqrt{\theta \Gamma}
        - \sqrt{\tau})\left(\sqrt{\theta \Gamma} + \frac{1}{ \lambda_{2} \sqrt{\tau}} \right)$,
    then $p_1^{opt} \in \mathcal{P}_2$. Otherwise, $p_1^{opt} \in \mathcal{P}_3$.
Here, $\tau:= \theta^{-1} \left( \lambda_{1}^{-1}  + \Gamma
\right) - \lambda_{2}^{-1}$, and $\Gamma$ and $\lambda_{i}$ are
defined in \eqref{eq:channelSNRandGamma}.
\end{proposition}

{\em Proof:} ~~See Appendix.

\vspace{0.5em}

\begin{figure}[h]
\centering \scalefig{0.5}\epsfbox{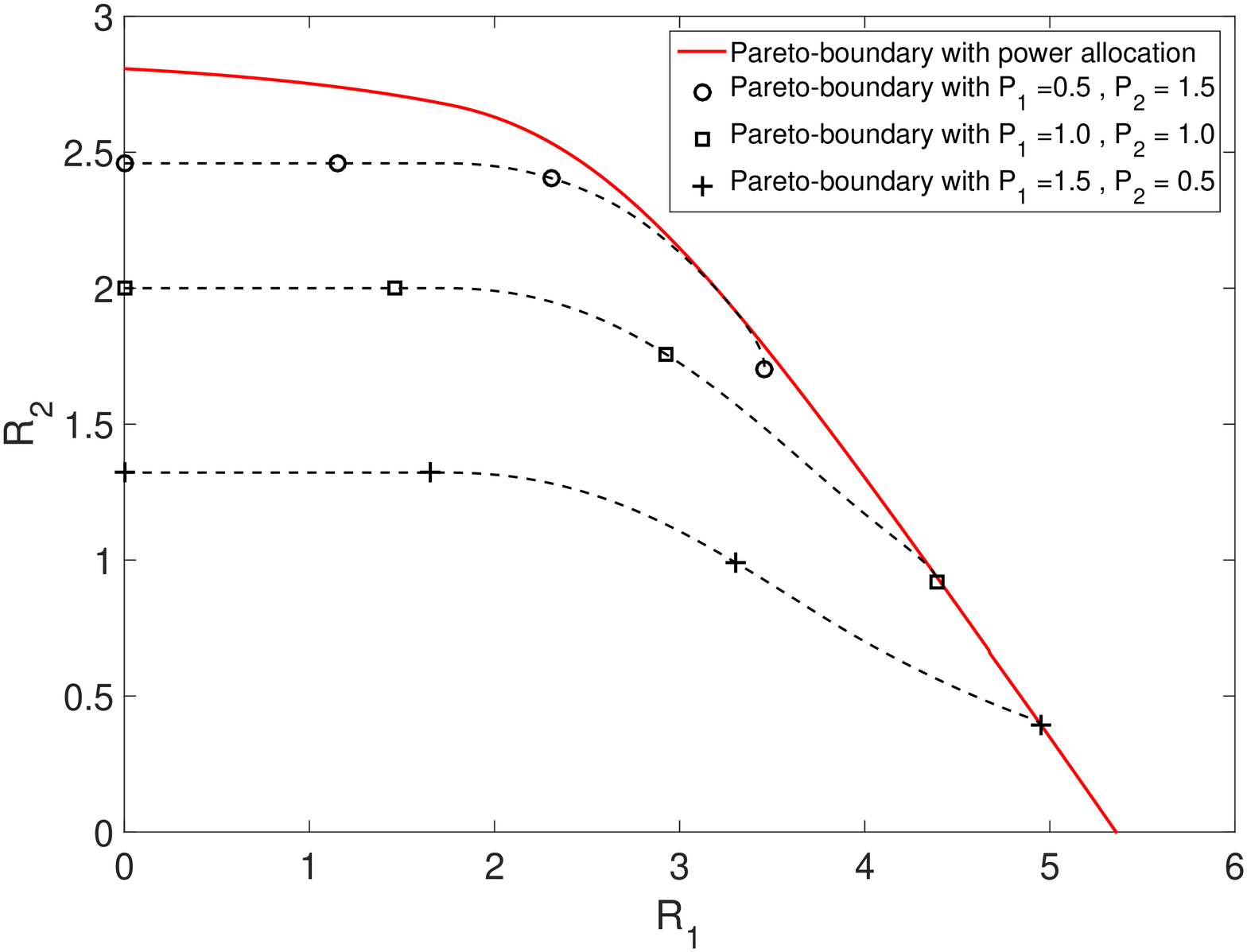}
\caption{Pareto-boundary: fixed power versus power allocation
($\|\hbf_1\|^2/\sigma_1^2=20, ~\|\hbf_2\|^2/\sigma_2^2=3,~\theta =
0.5$ and $P =2$)} \label{fig:PO_boudnary}
\end{figure}

Due to Proposition \ref{pro:PcregionDetermine} we know which of
the three cases in \eqref{eq:gamma2_p1} is applicable to the given
combination of $\hbf_1$, $\hbf_2$, $\sigma_1^2$, $\sigma_2^2$,
$P$, and $\gamma_1^*$. Once the set $\Pc_i$ to which $p_1^{opt}$
belongs is determined, optimal $p_1^{opt}$ can be found by
maximizing  the corresponding $\gamma_2^{*(i)}(p_1)$ in
\eqref{eq:gamma2_p1} with respect to $p_1$. A closed-form solution
from solving $\frac{d\gamma_2^{*(i)}(p_1)}{dp_1}=0$ seems
complicated but the solution can easily be found by a numerical
method.  The proposed algorithm to design Pareto-optimal beam
vectors and power allocation is summarized in Table
\ref{tb:algorithm01}.  The Pareto-boundary of a two-user MISO-NOMA
BC can  be computed by sweeping $R_1^*=\log(1+\gamma_1^*)$ and
computing the corresponding maximum $R_2^*=\log(1+\gamma_2^*)$. An
example is shown in Fig. \ref{fig:PO_boudnary}. It is seen that
power allocation significantly enlarges the achievable rate region
over the fixed-power beam-only design and optimal power allocation
is crucial for MISO-NOMA BC.

\begin{table}[t]
\caption{} \label{tb:algorithm01}
\begin{tabular}{p{450pt}}
\hline
\vspace{-0.5em}
\begin{itemize}
       \item[ ] \textbf{Pareto-optimal design for 2-user MISO-BC with SIC}
        \item[ ] \textbf{$[\sqrt{p_1}\wbf_1, \sqrt{p_2}\wbf_2] = \mathcal{D}(\hbf_1,\hbf_2,\sigma_1^2,\sigma_2^2,\gamma_1^*,P)$} \vspace{-1em}
\end{itemize}
\\ \hline \vspace{-0.8em}
\begin{itemize}
 \item[ ] \textbf{Input:} channel vectors $\hbf_1$, $\hbf_2$, noise power $\sigma_1^2, \sigma_2^2$, target SINR of user $1$ $\gamma_1^*$, and cluster total power $P$.

\item[] \textbf{Initialization:} $\lambda_{1} =
\|\hbf_1\|^2/\sigma_1^2$, $\lambda_{2} = \|\hbf_2\|^2/\sigma_2^2$,
$\theta = \frac{|\hbf_1^H\hbf_2|^2}{\|\hbf_1\|^2\|\hbf_2\|^2}$,
$\Gamma = \gamma_1^*/\lambda_{1}$, and $\tau =
\theta^{-1} \left( \lambda_{1}^{-1} +\Gamma \right) -
\lambda_{2}^{-1}$

    \begin{itemize}
        \item[] if $\theta \Gamma < \tau $

        \item[] \hspace{1em} obtain $p_1^{opt}$ maximizing $\gamma_2^{*(2)}$

        \item[] elseif $\tau \ge 0$ and  \item[]$~~~~~~~P \geq \Gamma + \frac{1}{1- \theta} (\sqrt{\theta \Gamma}
        - \sqrt{\tau})\left(\sqrt{\theta \Gamma} + \frac{1}{ \lambda_2\sqrt{\tau} } \right) $

        \item[] \hspace{1em} obtain $p_1^{opt}$ maximizing $\gamma_2^{*(2)}$

        \item[] else

        \item[] \hspace{1em} obtain $p_1^{opt}$ maximizing $\gamma_2^{*(3)}$

        \item[] endif

   \end{itemize}

   \item[] Obtain $\alpha_1^*, \beta_1^*$ and $\alpha_2^*$ using \eqref{eq:x_1_opt_with_p1}, \eqref{eq:PO_constraint1_3}  and $\eqref{eq:x_2_optimal_with_p1}$ with $p_1 = p_1^{opt}$, and
   obtain $\wbf_1$ and $\wbf_2$ from \eqref{eq:parameterization1} and
   \eqref{eq:parameterization2} with $\alpha_1^*,\beta_1^*$ and
   $\alpha_2^*$.
\item[] \textbf{Output:} $\sqrt{p_1^{opt}}\wbf_1$ and $\sqrt{P-p_1^{opt}}\wbf_2$
\end{itemize}
\\ \hline
\end{tabular}
\end{table}

\section{Two-user MISO broadcast channel with SIC: Performance Study}
\label{sec:perforamnce_analysis}

In the previous section, we developed a Pareto-optimal beam design
and power allocation algorithm for given channel vectors $\hbf_1$
and $\hbf_2$.   The performance of the Pareto-optimal design is a
function of the two (effective) channel vectors $\hbf_1$ and $\hbf_2$. In the
conventional ZF downlink beamforming with no SIC at the receivers,
two users with orthogonal channel vectors are preferred since
non-orthogonality between the two channel vectors reduces the
effective SINR of ZF beamforming\cite{Yoo&Goldsmith}. However, in
the considered  MISO-NOMA framework in which  user 1
intends to decode the interference from  user 2 and
subtracts it before decoding its own data whereas  user 2
treats the interference from user 1 as noise,
orthogonality between $\hbf_1$ and $\hbf_2$, i.e.,
 $\theta \approx 0$, and corresponding orthogonal beam vectors $\wbf_1$
and $\wbf_2$ (see \eqref{eq:parameterization1} and
\eqref{eq:parameterization2} with $\hbf_1 \perp \hbf_2$) do not
necessarily imply high
 performance.  Intuitively, if $\theta$ is small, user 2  receives
less interference from user 1, but user 1 has difficulty in
decoding the message of user 2 for SIC under the NOMA framework.
In this section, we investigate more on the two-user MISO-NOMA BC
before proceeding to overall user scheduling in the next section.

To gain some insight into good channel conditions for two-user
MISO-NOMA BCs, let us first consider the fixed power allocation
case with $p_1=p_2=1$ as described in Section
\ref{sec:fixed_power_allocation} and investigate the impact of the
angle $\theta$ between the two channel vectors when the magnitudes
are given. For given $||\hbf_1||$,  $||\hbf_2||$ and $\gamma_1^*$,
the SINR of user 2 $\gamma_2^*$ in \eqref{eq:gamma_2_fixed_power}
can be rewritten as a function of $\theta$ as
\begin{equation}
    \gamma_2^*(\theta) = \left\{ \begin{array}{ll}
                           \gamma_2^{*(1)} \; ( =  \frac{  {\lambda}_{1} }{1+ \Gamma {\lambda}_{1}}) & \quad \text{for case 1}, \\
                           \gamma_2^{*(2)}(\theta) \; ( = \frac{  {\lambda}_{1} }{1+ \Gamma {\lambda}_{1}} [\alpha_2^{*}(\theta)]^2)   &
                           \quad \text{for case 2}, \\
                           \gamma_2^{*(3)}(\theta) \; ( =\frac{ {\lambda}_{2} }{ {\lambda}_{2} [\alpha_1^{*}(\theta)]^2 + 1}) & \quad \text{for case 3},
                         \end{array}
       \right. \label{eq:gamma_2_fixed_power_eta}
\end{equation}
where
\begin{equation}
    \alpha_1^*(\theta) = \left\{ \begin{array}{cc}
                      0 & \quad   \text{if} \quad \Gamma \le 1 - \theta\\
                      \sqrt{\theta \Gamma} - \sqrt{(1-\theta) (1 - \Gamma )} & \quad  \text{if} \quad  \Gamma > 1 - \theta
                    \end{array}
    \right.
\end{equation}
and
\begin{equation}
    \alpha_2^*(\theta)  = \left\{ \begin{array}{ll}
                       1 & \quad \text{for case 1}\\
                       \frac{c(\theta)}{\sqrt{c^2(\theta) + (a-b(\theta))^2}} & \quad  \text{for case 2} \\
                       \sqrt{\theta}  & \quad \text{for case 3}.
                     \end{array}
     \right.
\end{equation}
Regarding optimal $\theta$ that maximizes $\gamma_2^*(\theta)$ in
\eqref{eq:gamma_2_fixed_power_eta}, we have the following
proposition:

\vspace{0.5em}

\begin{proposition} \label{pro:proposition1Angle}
Let $\lambda_1$,  $\lambda_2$ and $\Gamma$ be given. If
$\Gamma \in [\Gamma_1,~ \Gamma_2]$,
    where

    \begin{align}
       \Gamma_1 &= \frac{1}{2} \left(  1 +  \lambda_{2}^{-1}-\lambda_{1}^{-1} \right)
       - \frac{1}{2} \sqrt{( 1 + \lambda_{1}^{-1} + \lambda_{2}^{-1})^2 - 4\lambda_{2}^{-1}(1 + \lambda_{2}^{-1})} \nonumber \\
       \Gamma_2 &=  \frac{1}{2} \left(  1 +  \lambda_{2}^{-1}-\lambda_{1}^{-1} \right) + \frac{1}{2} \sqrt{( 1 + \lambda_{1}^{-1}
        + \lambda_{2}^{-1})^2 - 4\lambda_{2}^{-1}(1 +
       \lambda_{2}^{-1})}, \label{eq:Gamma1and2}
    \end{align}
 
     then optimal $\theta$ that maximizes $\gamma_2^{*}(\theta)$ in \eqref{eq:gamma_2_fixed_power_eta}
     is given by the region
  
     \begin{equation}
         \left\{ \theta | \theta_0 \leq \theta  \leq \frac{z_1 z_2 + 2\Gamma(1-\Gamma) + \sqrt{4\Gamma(1-\Gamma) [\Gamma(1-\Gamma) + z_1z_2 -z_2^2]}}{z_1^2 +
           4\Gamma(1-\Gamma)}\   \right\} \label{eq:opt_eta}
     \end{equation}
 
     where $z_1=\lambda_{1}^{-1} + 1-\Gamma$, $z_2 =\lambda_{2}^{-1} + 1-\Gamma$, and
     \begin{equation} \label{eq:theta_0}
        \theta_0 = \left\{~ \begin{array}{ll}
                              \frac{\lambda_{1}}{\lambda_{2}} \frac{1}{1 + \Gamma \lambda_{1}}, &~ \text{if }~~
                              \frac{\lambda_{1}}{\lambda_{2}} \frac{1}{1 + \Gamma \lambda_{1}} \leq 1 -\Gamma,  \\
                              \frac{z_1 z_2 + 2\Gamma(1-\Gamma) - \sqrt{4\Gamma(1-\Gamma) [\Gamma(1-\Gamma) + z_1z_2 -z_2^2]}}{z_1^2 + 4\Gamma(1-\Gamma)}, &~ \text{otherwise.}
                            \end{array}
          \right.
     \end{equation}
  If $\Gamma \notin [\Gamma_1,~\Gamma_2]$, optimal $\theta$ is   given by the region
    \begin{equation}
        \begin{array}{ll}
           \{\theta ~|~ \frac{\lambda_{2}}{\lambda_{1}} (1 + \Gamma \lambda_{1}) \leq \theta \leq 1-\Gamma\} &~~ \text{if }
           \Gamma \leq \frac{\lambda_{2}^{-1} - \lambda_{1}^{-1}}{1 + \lambda_{2}^{-1}} ~\mbox{and}~ \Gamma \notin [\Gamma_1,\Gamma_2],  \\
           \text{or}~\{\theta ~|~ \frac{\partial \gamma_2^{(2)}(\theta)}{\partial \theta} = 0\} &~~ \text{if }
           \Gamma > \frac{\lambda_{2}^{-1} - \lambda_{1}^{-1}}{1 + \lambda_{2}^{-1}}~\mbox{and}~ \Gamma \notin [\Gamma_1,\Gamma_2].
        \end{array}
     \end{equation}
\end{proposition}

{\em Proof:} ~~See Appendix.

\vspace{0.5em}

 From
Proposition \ref{pro:proposition1Angle} we obtain a more
insightful corollary as follows:

\vspace{0.5em}

\begin{corollary} \label{pro:corollaryAngle}
 Let $\lambda_1$,  $\lambda_2$ and $\Gamma$ be given.  When  $\lambda_1 = \lambda_2$,
    optimal $\theta$ for the 2-user MISO BC with SIC  is given by  the set $\{\theta ~|~ \theta_0 \le \theta \le 1\}$ with  $\theta_0$ reduced to
     \begin{equation}
        \theta_0 = \left\{~ \begin{array}{ll}
                              \frac{1}{1 + \Gamma \lambda_{1}}  ~~& \text{ if }~~
                               \frac{1}{1 + \Gamma \lambda_{1}} \leq 1 -\Gamma \\
                              \frac{z_1^2}{z_1^2 + 4\Gamma(1-\Gamma)} ~~& \text{ if }~~   \frac{1}{1 + \Gamma \lambda_{1}}  > 1 -\Gamma
                            \end{array}
          \right. .
     \end{equation}
\end{corollary}

{\em Proof:} ~~See Appendix.

\vspace{0.5em}

\noindent Corollary  \ref{pro:corollaryAngle} states that in two-user
MISO BCs with SIC with the same channel magnitudes
$\lambda_1=\lambda_2$ and the same power $p_1=p_2=1$, two
aligned channel vectors are preferred to two orthogonal channels
and channel alignment beyond a certain angle is all optimal.

Although Proposition \ref{pro:proposition1Angle} and Corollary
\ref{pro:corollaryAngle} provide some insight into good channel
conditions in the SIC BC case, the assumptions for Proposition
\ref{pro:proposition1Angle} and Corollary \ref{pro:corollaryAngle}
are not valid in the actual NOMA situation in which power allocation is
applied. Unfortunately, the optimal power  $p_1^{opt}$ was not
obtained in closed form in the previous section and this puts
difficulty on analysis of the impact of channel angle on the
performance.  Hence, in the actual case, to enable analysis we derive the SINR
$\gamma_2^*$ for user 2 as a function of $\theta$ by
assuming the simple power allocation method that assigns minimum
power $p_{1,min}(=\gamma_1^*/\lambda_1=\Gamma)$
 to
achieve the target SINR $\gamma_1^*$ to user $1$ and assigns the
rest of power $P$  to user $2$.  The simple power allocation
strategy is based on the assumption that user $1$ has a strong
channel and is limited by channel's DoF such as bandwidth, whereas  user $2$ with  a
weak channel is limited by noise and needs to receive more power. For this simple power allocation method,  the
optimal $\gamma_2^{*}$ is obtained by substituting
$p_1=p_{1,min}=\Gamma$ into \eqref{eq:gamma2_p1}
  and given after some manipulation  in closed form as

\begin{equation} \label{eq:gamma2_simple}
    \gamma_2^* =\left\{ \begin{array}{ll}
                          \frac{P-\Gamma}{\Gamma} \frac{1}{\lambda_1^{-1} \Gamma^{-1} + 1} \left[1+ \frac{\theta}{1-\theta}\left(\sqrt{\frac{1+\lambda_2^{-1} \Gamma^{-1} \theta^{-1}  }{1+ \lambda_1^{-1} \Gamma^{-1}}} - 1\right)^2\right]^{-1}, &\quad \text{~if } \theta \leq \theta_1 \\
                          \frac{P-\Gamma}{\Gamma} \frac{1}{\theta +\lambda_2^{-1} \Gamma^{-1}}, &\quad \text{~if } \theta > \theta_1,
                        \end{array}
    \right.
\end{equation}

\begin{figure*}[!t]
\centerline{
\SetLabels
\L(0.16*-0.1) (a) \\
\L(0.49*-0.1) (b) \\
\L(0.82*-0.1) (c) \\
\endSetLabels
\leavevmode
\strut\AffixLabels{
\scalefig{0.33}\epsfbox{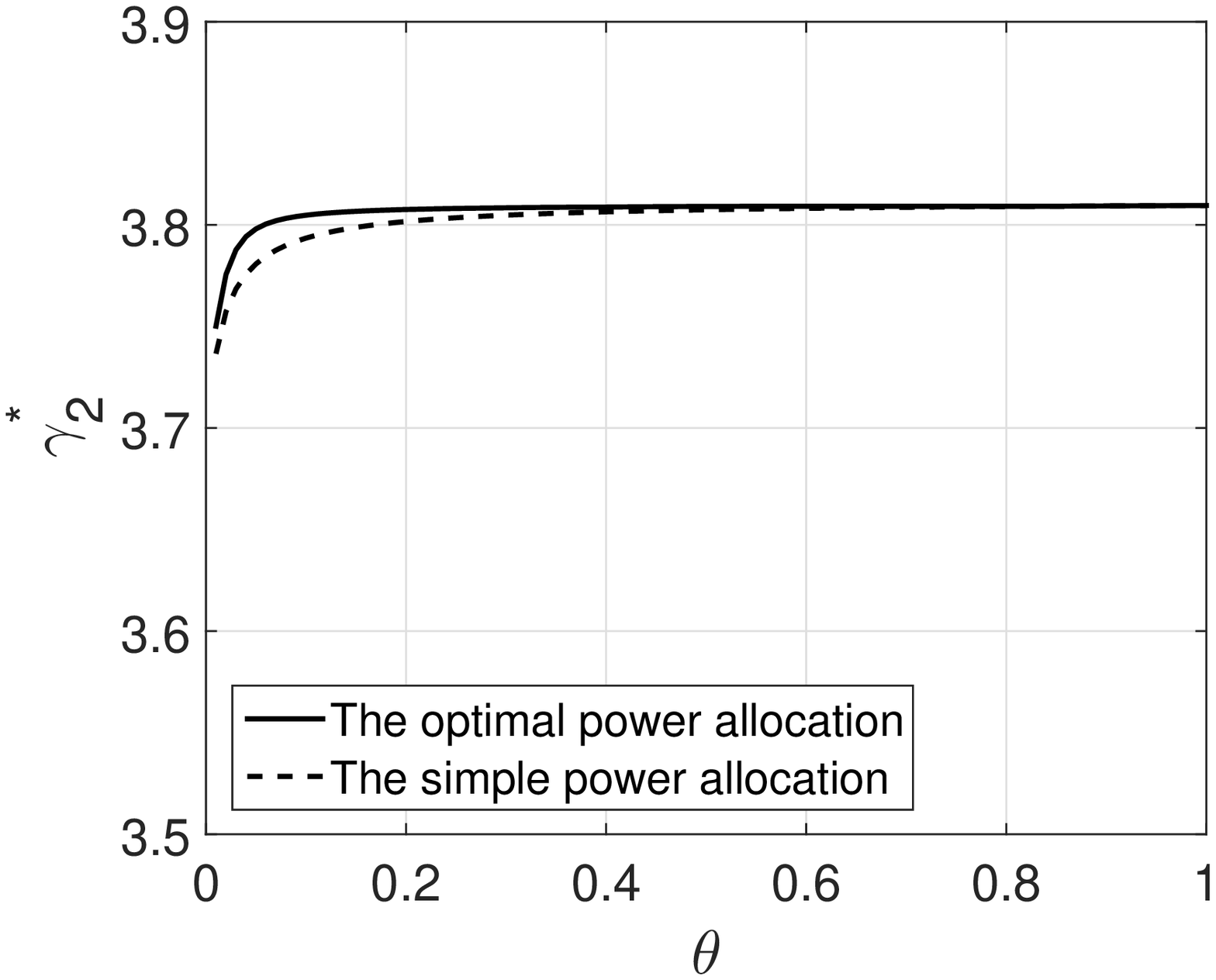}
\scalefig{0.33}\epsfbox{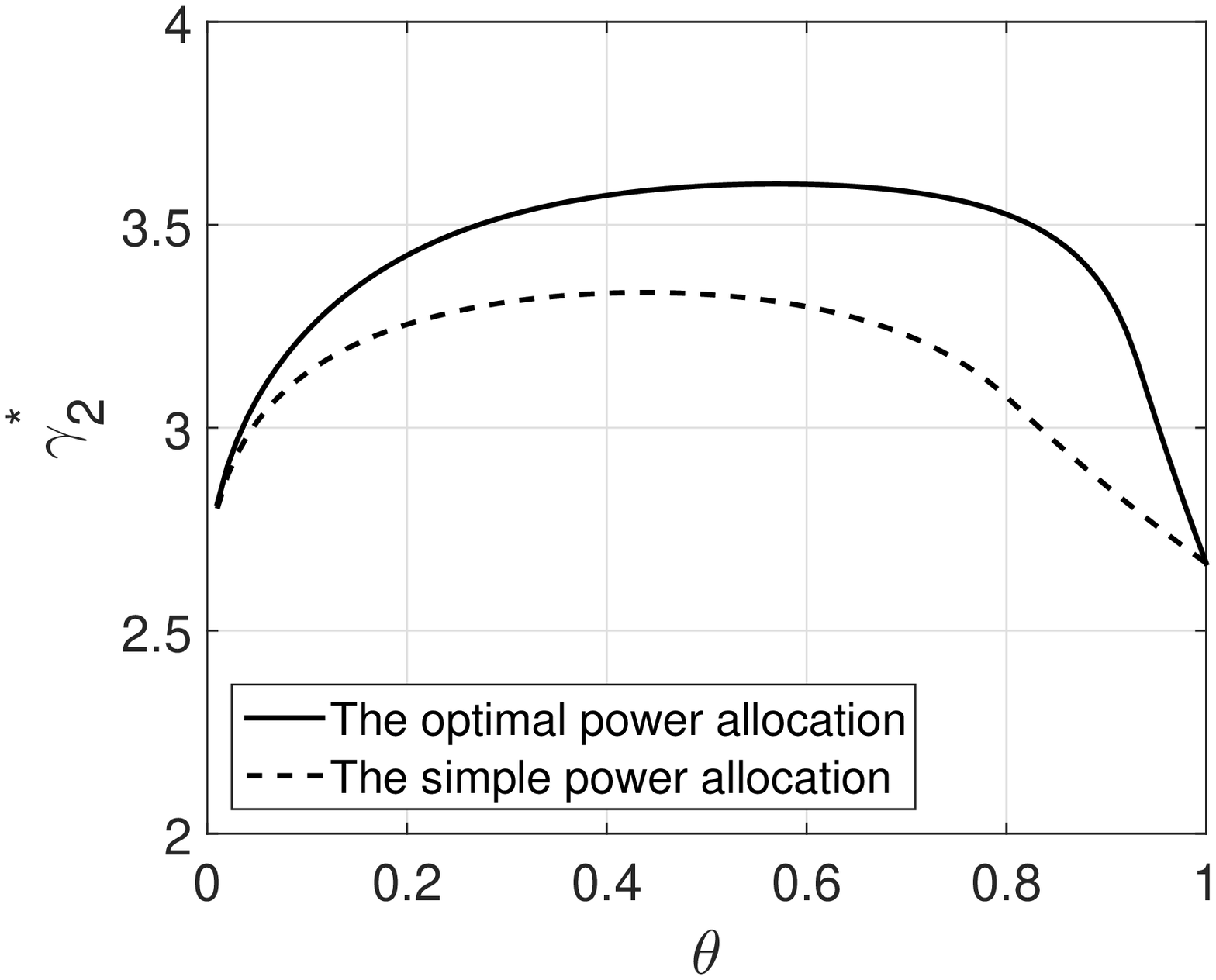}
\scalefig{0.33}\epsfbox{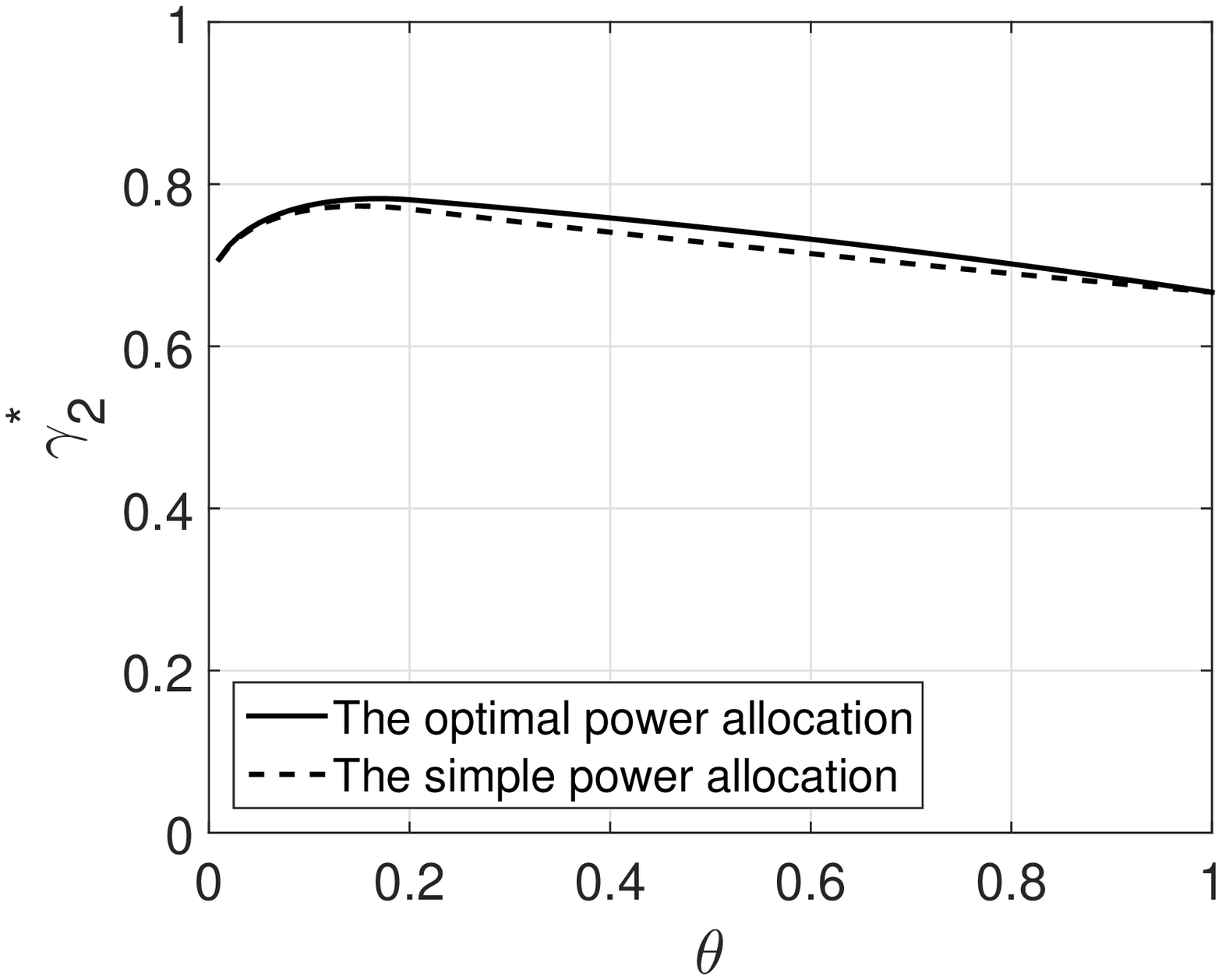} } }
\vspace{1.5em} \caption{$\gamma_2^*$ in \eqref{eq:gamma2_simple} as a function of $\theta$: (a) $\lambda_1= 10$, $\lambda_2= 10$, $\Gamma= 2$, $P=10$ , (b) $\lambda_1= 10$, $\lambda_2= 1$, $\Gamma= 2$, $P=10$, and (c) $\lambda_1= 10$, $\lambda_2= 0.1$, $\Gamma= 2$, $P=10$.} \label{fig:anglePlot}
\end{figure*}

\noindent where  $\theta_1 := \frac{1}{2}\left[-\lambda_2^{-1}\Gamma^{-1} +
\sqrt{\lambda_2^{-2} \Gamma^{-2} + 4 (\lambda_1^{-1} \Gamma^{-1} +
1)}\right]$.

Examples of $\gamma_2^*$ in \eqref{eq:gamma2_simple}
as a function of $\theta$  are shown in Fig. \ref{fig:anglePlot} together with the optimal $\gamma_2^*$ obtained by running the algorithm in Table I.
It is seen that the $\gamma_2^*$ behavior depends on the relative magnitude of $\lambda_1$ and $\lambda_2$ through the two performance limiting  factors: the SIC processing at user 1 and the SINR of user 2, as seen in \eqref{eq:PO_constraint1_2}.
In the case of $\lambda_1=\lambda_2 \gg 0$,  the
performance of user 2 is not limited by noise at user 2 but is limited by signal-to-interference ratio (SIR). Thus, in this case it is preferred that channel vectors are aligned and more power is allocated to user 2 under the constraint that the required SINR for user 1 is satisfied. By doing so, SIC at user 1 is easy and SIR at user 2 is high. This behavior is evident in Fig. \ref{fig:anglePlot}(a). It is seen that the optimal power control and the simple power allocation strategy yield similar performance in Fig. \ref{fig:anglePlot}(a). (The
behavior for $\lambda_1=\lambda_2$ with power control seems similar to that stated in
Corollary \ref{pro:corollaryAngle}.)
However, in the medium asymmetric case of $\lambda_1 > \lambda_2$ as in Fig. \ref{fig:anglePlot}(b), there is a trade-off between the two performance limiting factors. When two channels are orthogonal, SIC at user 1 is difficult. On the other hand, when two channels are aligned, interference from user 1 to user 2 at user 2 is high. Hence, the performance is good when the two user channels are neither too orthogonal  nor too aligned. This behavior is evident in Fig. \ref{fig:anglePlot}(b). Note that in this case, there is a large gap between the optimal power control and the simple power allocation.

In addition to the above simple power allocation result, we have
another result exploiting the fact $\lambda_1 \gg \lambda_2$ in
actual NOMA, given by the following proposition:
\begin{proposition}  \label{prop:l1ggl2}
    For given $\lambda_1$ and $\gamma_1^*$, if $\theta \ne 0$,  as $\lambda_2 \rightarrow 0$, the optimal
    power coefficient  $p_1^{opt} \rightarrow p_{1,min}=\Gamma$;  the corresponding $\gamma_2^*$ converges to
        $\gamma_2^* = \frac{P-\Gamma}{\Gamma} \frac{1}{\theta + \lambda_2^{-1} \Gamma^{-1}}$;
    and the beam vectors converge to
    $\sqrt{p_1}\wbf_1 = \sqrt{\Gamma} \frac{\hbf_1}{\|\hbf_1\|}$ and $\sqrt{p_2}\wbf_2 = \sqrt{P-\Gamma} \frac{\hbf_2}{\|\hbf_2\|}$.
    That is, both users use matched-filtering beams.
\end{proposition}

{\em Proof:} ~~See Appendix.

Note that $\gamma_2^*$ in Proposition \ref{prop:l1ggl2} coincides with the second formula in \eqref{eq:gamma2_simple}. This is because $
\theta_1$ in \eqref{eq:gamma2_simple} converges to $0^+$ as $
\lambda_2 \rightarrow 0$ for given $\lambda_1$ and $\Gamma$, and the second formula in \eqref{eq:gamma2_simple} is valid in this case.  By  Proposition \ref{prop:l1ggl2}, if $\lambda_2$ is sufficiently small compared to $\lambda_1$, matched filtering beams for both users with minimum power to user 1 satisfying the target SINR are optimal regardless of the angle between the two channel vectors.  This is because if $\lambda_1 \gg \lambda_2$, the limitation for $\gamma_2^*$ results from the SINR of user 2 at user 2. Hence, maximum power should be delivered to user 2 with matched filtering beam $\wbf_2$ by assigning minimum power to user 1 with matched filtering beam $\wbf_1$.
This behavior is evident in Fig. \ref{fig:anglePlot}(c).
It is seen in Fig. \ref{fig:anglePlot}(c) that the simple power allocation method almost achieves the optimal performance for all $\theta$.

\section{The Proposed Scheduling Method for $K$-User MISO-NOMA Downlink}
\label{sec:scheduling}

\begin{figure*}[t!]
\centerline{
 \begin{psfrags}
  \psfrag{h1}[c]{\small $\tilde{\hbf}_{1}^{(1)}$} %
  \psfrag{h2}[c]{\small $\tilde{\hbf}_{1}^{(2)}$} %
  \psfrag{w1}[c]{\small $\wbf_{1}^{(1)}$} %
  \psfrag{w2}[c]{\small $\wbf_{2}^{(1)}$} %
  \psfrag{h3}[l]{\small $\tilde{\hbf}_{1}^{(3)}$} %
  \psfrag{g1}[l]{\small $\tilde{\hbf}_{2}^{(1)}$} %
  \psfrag{ch1}[c]{\small $\Lc^\perp(\tilde{\hbf}_{1}^{(2)})$} %
  \psfrag{ch12}[c]{\small $\Lc^\perp([\tilde{\hbf}_{1}^{(2)},\tilde{\hbf}_{1}^{(3)}])$} %
  \psfrag{ph1}[r]{\tiny $\Pibf^\perp_{\tilde{\hbf}_{1}^{(2)}}\tilde{\hbf}_{1}^{(1)}$} %
  \psfrag{pg1}[l]{\tiny $\Pibf^\perp_{\tilde{\hbf}_{1}^{(2)}}\tilde{\hbf}_{1}^{(2)}$} %
  \psfrag{pph1}[l]{\tiny $\Pibf^\perp_{[\tilde{\hbf}_{1}^{(2)},\tilde{\hbf}_{1}^{(3)}]}\tilde{\hbf}_{1}^{(1)}$} %
  \psfrag{ppg1}[l]{\tiny $\Pibf^\perp_{[\tilde{\hbf}_{1}^{(2)},\tilde{\hbf}_{1}^{(3)}]}\tilde{\hbf}_{1}^{(2)}$} %
  \psfrag{b}[c]{\small (a)} %
  \psfrag{a}[c]{\small (b)} %
    \scalefig{0.85}\epsfbox{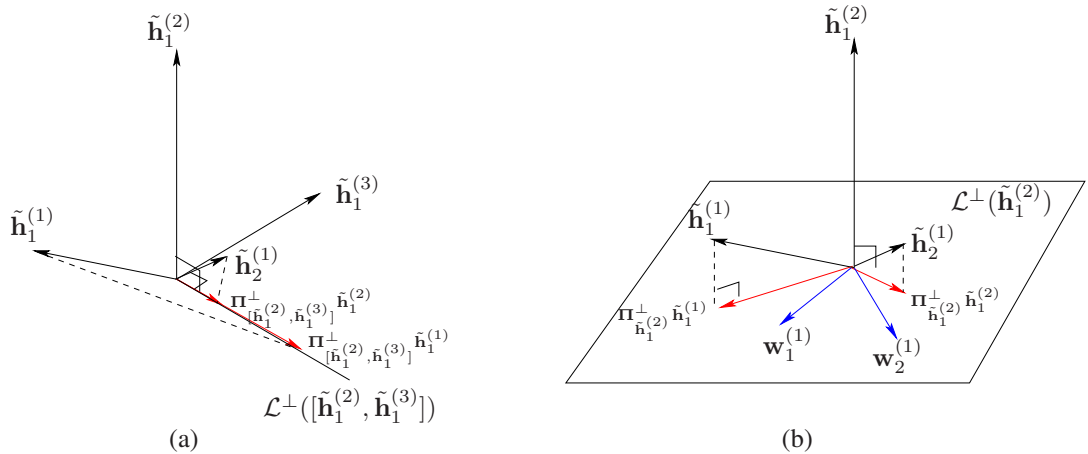}
\end{psfrags}
} \caption{Beam design in  the example of $N_t=3$: (a) $K_c=N_t$ and (b) $K_c=N_t -1$}
\label{fig:overalldesignfigure}
\end{figure*}

Now,  we  propose our overall user scheduling/pairing and beam design method for the $K$-user MISO-NOMA downlink with $N_t$ BS antennas based on
 the results in the previous sections.
In  MISO-NOMA downlink scheduling and beam design,   two major
aspects should be taken into account simultaneously to guarantee
 good system performance: One is controlling ICI to reduce
interference from other clusters and the other is pairing and beam
design for the paired users in each cluster for maximum
performance.  Recall that the gain of NOMA lies in the case that
strong users are in the DoF(such as bandwidth)-limited regime and weak users are limited by
noise\cite[P. 239]{Tse:book}. In MU-MISO NOMA, the beneficial situation
for NOMA should be maintained, i.e., the high channel quality for
strong users should be maintained by SIC and proper interference
control, and low SINRs of  weak users should be leveraged by
assigning more power to weak users. To be consistent with this
design principle, {\em  ICI should be eliminated for strong users}
by proper measures.
With these considerations, we propose the
following user scheduling, pairing and beam design method for
$K$-user MISO-NOMA downlink composed of two steps under the
assumption that all channel vector information  is available at the BS and the
thermal noise variance is known.

\vspace{0.5em}

{\em Algorithm 1: Overall User Scheduling and Beam Design}

\vspace{0.5em}

\noindent \textbf{Step 1:} In the first step, we  run the semi-orthogonal user
       selection (SUS) algorithm \cite{Yoo&Goldsmith} targeting selection of $N_t$ users from the strong user set $\Kc_1$.
Then, the SUS algorithm returns $K_c (\le N_t)$ users with roughly orthogonal
channel vectors out of the $K/2$ users in $\Kc_1$. (Depending on
the size $|\Kc_1|$ and the semi-orthogonality parameter, the SUS
algorithm  may return  users  less than $N_t$ especially for large $N_t$ although we target selecting $N_t$ users\cite{Yoo&Goldsmith}.) We set
these $K_c$ users returned by the SUS algorithm as the $K_c$
strong users in $K_c$ clusters  (one user
for each cluster). Let their actual channel vectors be
$\tilde{\hbf}_{1}^{(1)},\cdots,\tilde{\hbf}_{1}^{(K_c)}$.

\vspace{1em}

\noindent \textbf{Step 2:} Weak user selection and overall beam design

\noindent \textbf{Initialization:}   $\Gamma$ is given.
    \begin{align}
     \Kc_2 &= \{ 1, \ldots , K/2 \} ~\text{(the original weak user set)}, \\
      \Sc_1 &: \text{the set of selected strong users from step 1},\\
      \Sc_2 &\leftarrow \phi ~~\text{(the set of selected weak users)}, \\
      \widehat{\Wbf} &= \left[\frac{\Pibf_{\tilde{\Hbf}_1}^\perp \tilde{\hbf}_{1}^{(1)}}{||\Pibf_{\tilde{\Hbf}_1}^\perp \tilde{\hbf}_{1}^{(1)}||}, \cdots, \frac{\Pibf_{\tilde{\Hbf}_{K_c}}^\perp \tilde{\hbf}_{1}^{(K_c)}}{||\Pibf_{\tilde{\Hbf}_{K_c}}^\perp \tilde{\hbf}_{1}^{(K_c)}||} \right], \\
      \widetilde{\Wbf}_1 &= [~] \quad \text{and} \quad \widetilde{\Wbf}_2 = [~],\\
      k &= 1.
    \end{align}

\noindent \textbf{Iteration:}
    \begin{itemize}
        \item[] \textbf{while $k \leq K_c$ do}

            \begin{itemize}
                \item[]S.1:              For each user $u \in \Kc_2$, estimate ICI plus AWGN:
                \begin{eqnarray}
                    \hat{\sigma}_u^2 &=& \sum_{l < k}  (|\tilde{\gbf}_u^H \widetilde{\Wbf}_{1}(l)|^2 + |\tilde{\gbf}_u^H \widetilde{\Wbf}_{2}(l)|^2) + \sum_{l > k} P |\tilde{\gbf}_u^H \widehat{\Wbf}(l)|^2 + \epsilon_u^2,  \label{eq:estICIwgn}
                \end{eqnarray}
                where $\widehat{\Wbf}(l), \widetilde{\Wbf}_{1}(l)$ and $\widetilde{\Wbf}_{2}(l)$ are the $l$-th columns of $\widehat{\Wbf}, \widetilde{\Wbf}_1$ and $\widetilde{\Wbf}_2$ respectively, and $\tilde{\gbf}
                _u$ is the actual channel vector of user $u$ in $\Kc_2$.
                                With obtained $\hat{\sigma}_u^2$ and given $\Gamma$, compute the maximum SINR $\gamma_{2}^*(u)$ of user $u$ when user $u$ is paired with the channel $\tilde{\hbf}_1^{(k)}$ of user 1 of cluster $k$, as described in Section \ref{sec:PA_problem} or   Section IV  by setting $\lambda_1 = \frac{\|\Pibf_{\tilde{\Hbf}_k}^\perp \tilde{\hbf}_{1}^{(k)}\|^2}{\sigma^2_1}$, $\lambda_2 = \frac{\|\Pibf_{\tilde{\Hbf}_k}^\perp \tilde{\gbf}_{u}\|^2}{\hat{\sigma}^2_u}$, and  $\theta  = \frac{\left|(\Pibf_{\tilde{\Hbf}_k}^\perp \tilde{\hbf}_{1}^{(k)})^H(\Pibf_{\tilde{\Hbf}_k}^\perp \tilde{\gbf}_{u}) \right|^2}{\|\Pibf_{\tilde{\Hbf}_k}^\perp \tilde{\hbf}_{1}^{(k)}\|^2\|\Pibf_{\tilde{\Hbf}_k}^\perp \tilde{\gbf}_{u}\|^2}$.
                \item[]S.2: Select the weak user of cluster $k$ as follows:
                        \begin{align}
                            u^* &= \mathop{\arg\max}\limits_{u \in \Kc_2} \gamma_{2}^*(u) \\
                            \Sc_2 &\leftarrow \Sc_2 \cup \{u^*\} \\
                            \tilde{\hbf}_2^{(k)}&= \tilde{\gbf}_{u^*}
                            \end{align}
                        and design  $[\sqrt{p_1}\tilde{\wbf}_{1}^{(k)}, \sqrt{p_2}\tilde{\wbf}_2^{(k)}] = \mathcal{D}(\Pibf_{\tilde{\Hbf}_k}^\perp \tilde{\hbf}_{1}^{(k)},$ $\Pibf_{\tilde{\Hbf}_k}^\perp \tilde{\hbf}_2^{(k)},\sigma_1^2,\hat{\sigma}^2_{u^*},\lambda_1\Gamma,P)$
                         by using the two-user Pareto-optimal design algorithm in Table. \ref{tb:algorithm01}.
                \item[]S.3: Store the designed beams, remove $u^*$ in $\Kc_2$, and repeat until $k = K_c$:
                    \begin{align*}
                          \widetilde{\Wbf}_1&\leftarrow [\widetilde{\Wbf}_1, \sqrt{p_1}\tilde{\wbf}_1^{(k)}], ~~\widetilde{\Wbf}_2\leftarrow [\widetilde{\Wbf}_2, \sqrt{p_2}\tilde{\wbf}_2^{(k)}] \\
                        \Kc_2 &\leftarrow \Kc_2 \setminus \{u^*\} \\
                        k &\leftarrow k + 1
                    \end{align*}
            \end{itemize}
        \item[] \textbf{end while}
   \end{itemize}

\begin{remark}
Note that the computation of $\gamma_2^*(u)$ and the Pareto-optimal beam design for two users in each cluster in steps S.1 and S.2 in the while loop of Step 2 of Algorithm 1  is based on the {\em projected effective channels}. Note that the projected effective channels $\Pibf_{\tilde{\Hbf}_k}^\perp \tilde{\hbf}_{1}^{(k)}$ and $\Pibf_{\tilde{\Hbf}_k}^\perp \tilde{\hbf}_2^{(k)}$ lie in $\Lc^\perp(\tilde{\Hbf}_k)$. Thus, the corresponding Pareto-optimal beams $\wbf_1^{(k)}$ and $\wbf_2^{(k)}$ lie in $\Lc^\perp(\tilde{\Hbf}_k)$ by the property of Pareto-optimal beams (see \eqref{eq:parameterization1}
and \eqref{eq:parameterization2})\cite{Jorswieck}, and hence $\wbf_i^{(k)} = \Pibf^\bot_{\tilde{\Hbf}_k}\wbf_i^{(k)}= \tilde{\wbf}_i^{(k)}, ~i=1,2$ for \eqref{eq:systemmodelproj}.  Hence, there exists no ICI to all strong users with the proposed beam design.
\end{remark}

\begin{remark}
If $K_c= N_t$, then $\tilde{\Hbf}_k$ has nullity of one, and
$\Pibf_{\tilde{\Hbf}_k}^\perp\tilde{\hbf}_{1}^{(k)}$ and
$\Pibf_{\tilde{\Hbf}_k}^\perp\tilde{\hbf}_{2}^{(k)}$ are aligned, as shown in
Fig. \ref{fig:overalldesignfigure}(a). In this case, only
Pareto-optimal power control is applied for each cluster by the proposed design method. (Note that the algorithm in
Table \ref{tb:algorithm01} is still applicable in case of two
aligned input channel vectors.)  On the other hand,
 if the number $K_c$ of the
returned users by the SUS algorithm in Step 1 is less than $N_t$ (which is often true for large $N_t$ with small $K$\cite{Yoo&Goldsmith}), then $\tilde{\Hbf}_k$ has nullity larger than or equal to two. In this case,
the projected effective channels
$\Pibf_{\tilde{\Hbf}_k}^\perp \tilde{\hbf}_{1}^{(k)}$ and
$\Pibf_{\tilde{\Hbf}_k}^\perp \tilde{\hbf}_{2}^{(k)}$ span a 2-dimensional
(2-D) space and the full 2-D Pareto-optimal beam design is applicable for each
cluster, as shown in
Fig. \ref{fig:overalldesignfigure}(b).  This is another advantage of the proposed method over the previous methods \cite{Kim13:MILCOM,Liu15:ICCW,Sayed17:WD} based on simple spatial ZF beam design ignoring the case of $K_c < N_t$.
\end{remark}

\begin{remark}
In the step \eqref{eq:estICIwgn} of computation of ICI and AWGN for each candidate weak user for cluster $k$, already designed beam vectors are used up to cluster $k-1$ and the beam estimates
$\sqrt{P}\frac{\Pibf_{\tilde{\Hbf}_{1}}^\perp \tilde{\hbf}_{1}^{(k+1)}}{||\Pibf_{\tilde{\Hbf}_{k+1}}^\perp \tilde{\hbf}_{1}^{(k+1)}||}, \cdots,$

$\sqrt{P}\frac{\Pibf_{\tilde{\Hbf}_{K_c}}^\perp \tilde{\hbf}_{1}^{(K_c)}}{||\Pibf_{\tilde{\Hbf}_{K_c}}^\perp \tilde{\hbf}_{1}^{(K_c)}||}$ are used for undesigned clusters $k+1,\cdots,K_c$.
\end{remark}

\begin{remark}
 Note
that there is length reduction from the actual channel $\tilde{\hbf}_{1}^{(k)}$ to the effective channel
$\Pibf_{\tilde{\Hbf}_k}\tilde{\hbf}_{1}^{(k)}$ of each strong user. This reduction is the
typical effective gain loss associated ZF,
but the loss is not significant because the strong channels
$\tilde{\hbf}_{1}^{(1)},\cdots, \tilde{\hbf}_{1}^{(K_c)}$ are semi-orthogonal by
the SUS algorithm\cite{Yoo&Goldsmith}. Only the weak users
experience ICI whereas ICI to the strong users is completely
removed in the proposed method to be consistent with the NOMA
design principle. However, the weak users are selected by
considering all the factors, i.e., the ICI, projection onto
$\Lc^\perp(\tilde{\Hbf}_k)$ and the friendliness with the strong
users to yield good performance.
\end{remark}

\section{Numerical results}
\label{sec:NumericalResult}

In this section, we provide some numerical results to evaluate the
performance of the proposed scheduling,  beam
design and power allocation method described in Section  \ref{sec:scheduling}
for MU-MISO-NOMA downlink.

\begin{figure*}[t]
\centerline{
\SetLabels
\L(0.25*-0.1) (a) \\
\L(0.75*-0.1) (b) \\
\endSetLabels
\leavevmode
\strut\AffixLabels{
\scalefig{0.5}\epsfbox{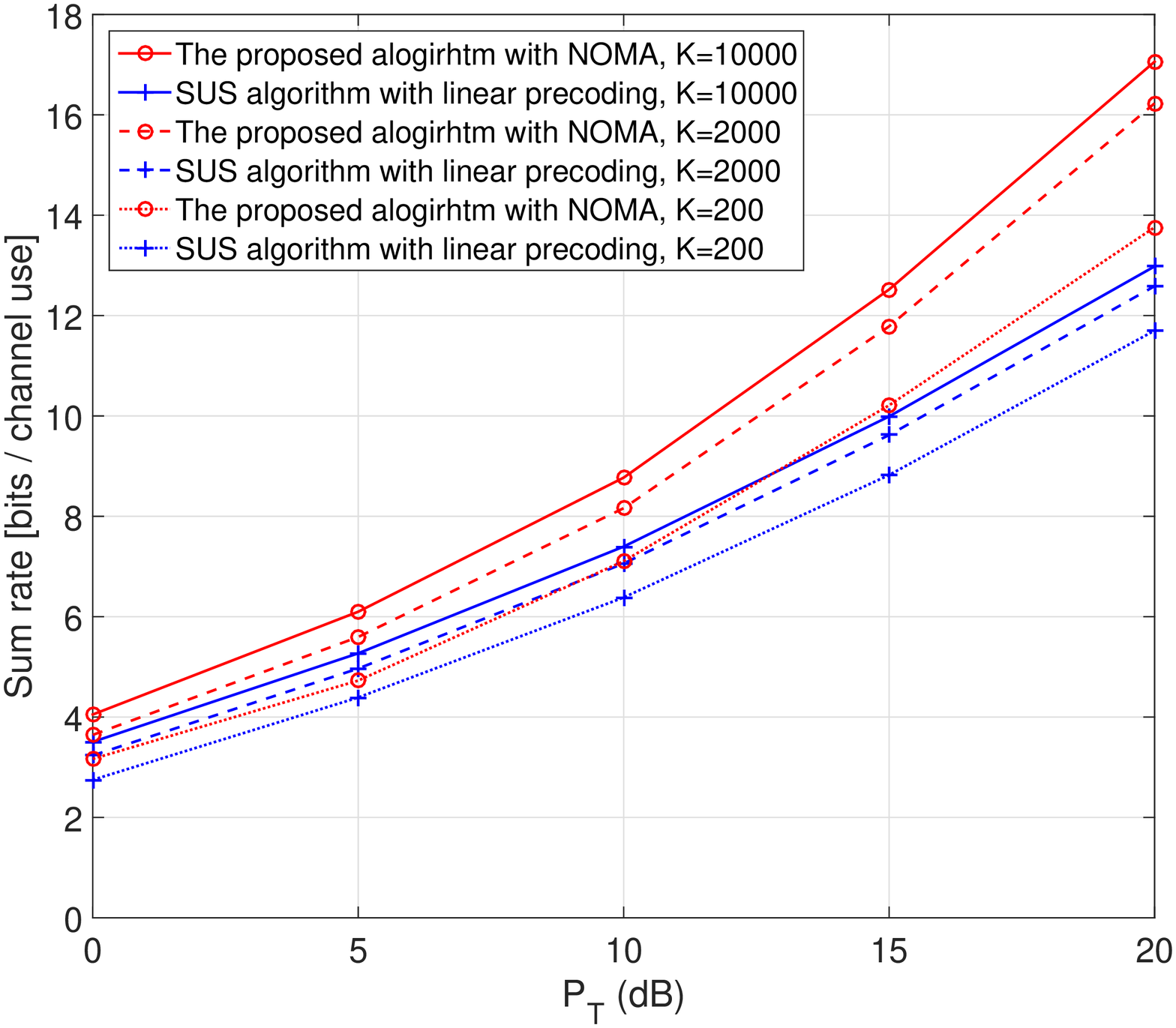}
\scalefig{0.5}\epsfbox{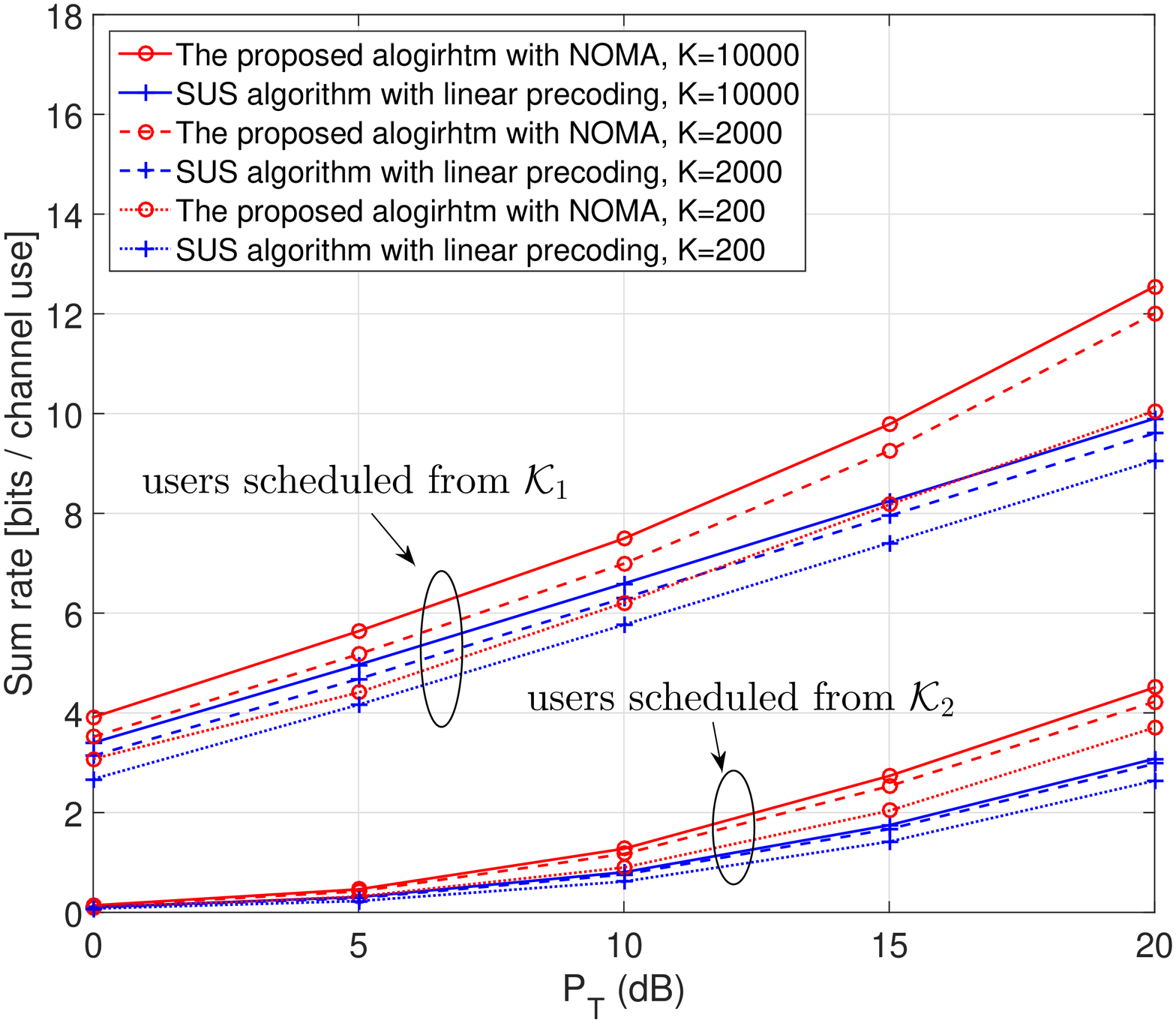} } }
\vspace{1.5em} \caption{Sum rate ($N_t = 2$) : (a) total sum rate and
(b) separate sum rates from $\Kc_1$ and from $\Kc_2$} \label{fig:sumrate_Nt2}
\end{figure*}

\begin{figure*}[!t]
\centerline{
\SetLabels
\L(0.25*-0.1) (a) \\
\L(0.75*-0.1) (b) \\
\endSetLabels
\leavevmode
\strut\AffixLabels{
\scalefig{0.5}\epsfbox{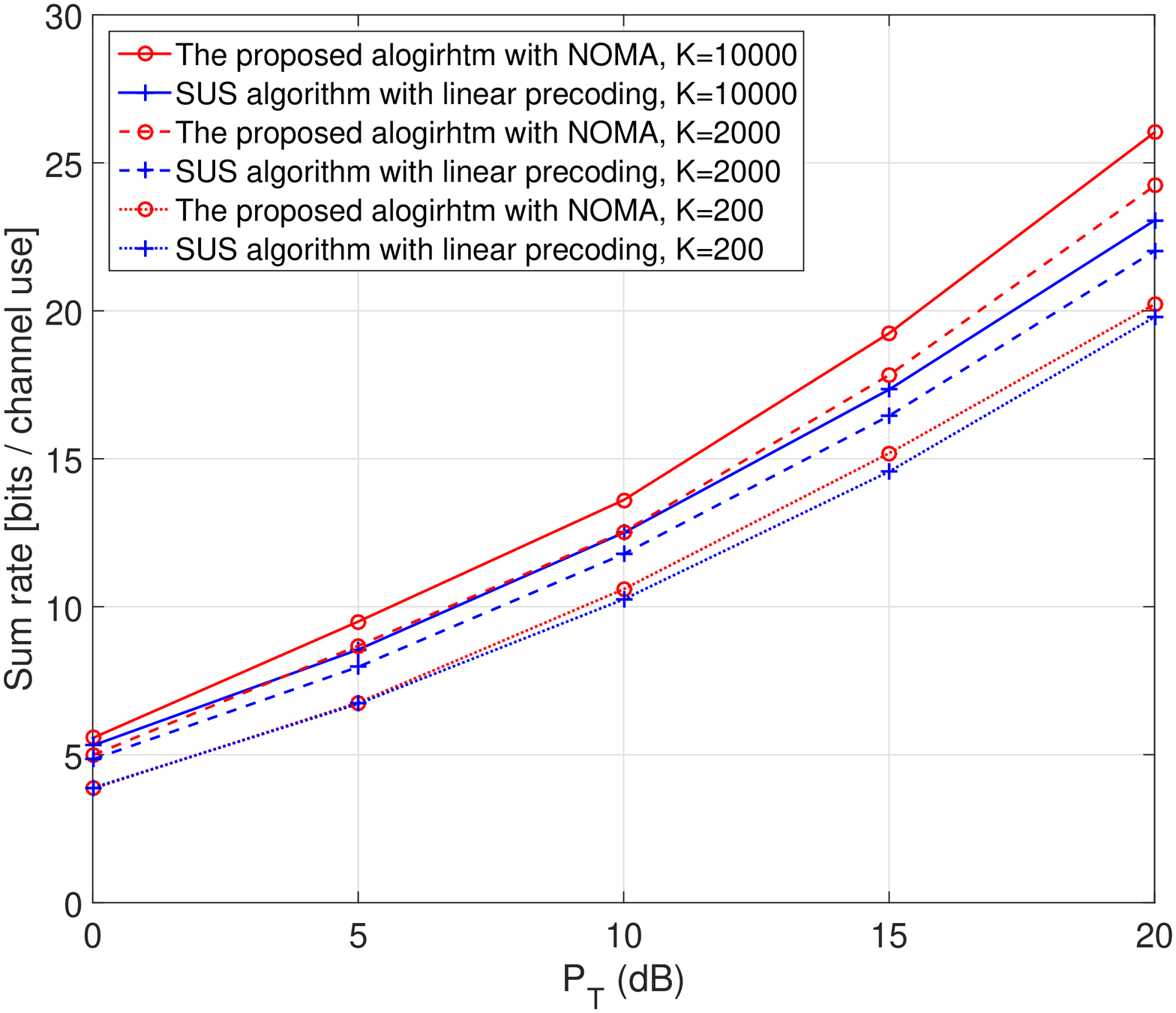}
\scalefig{0.5}\epsfbox{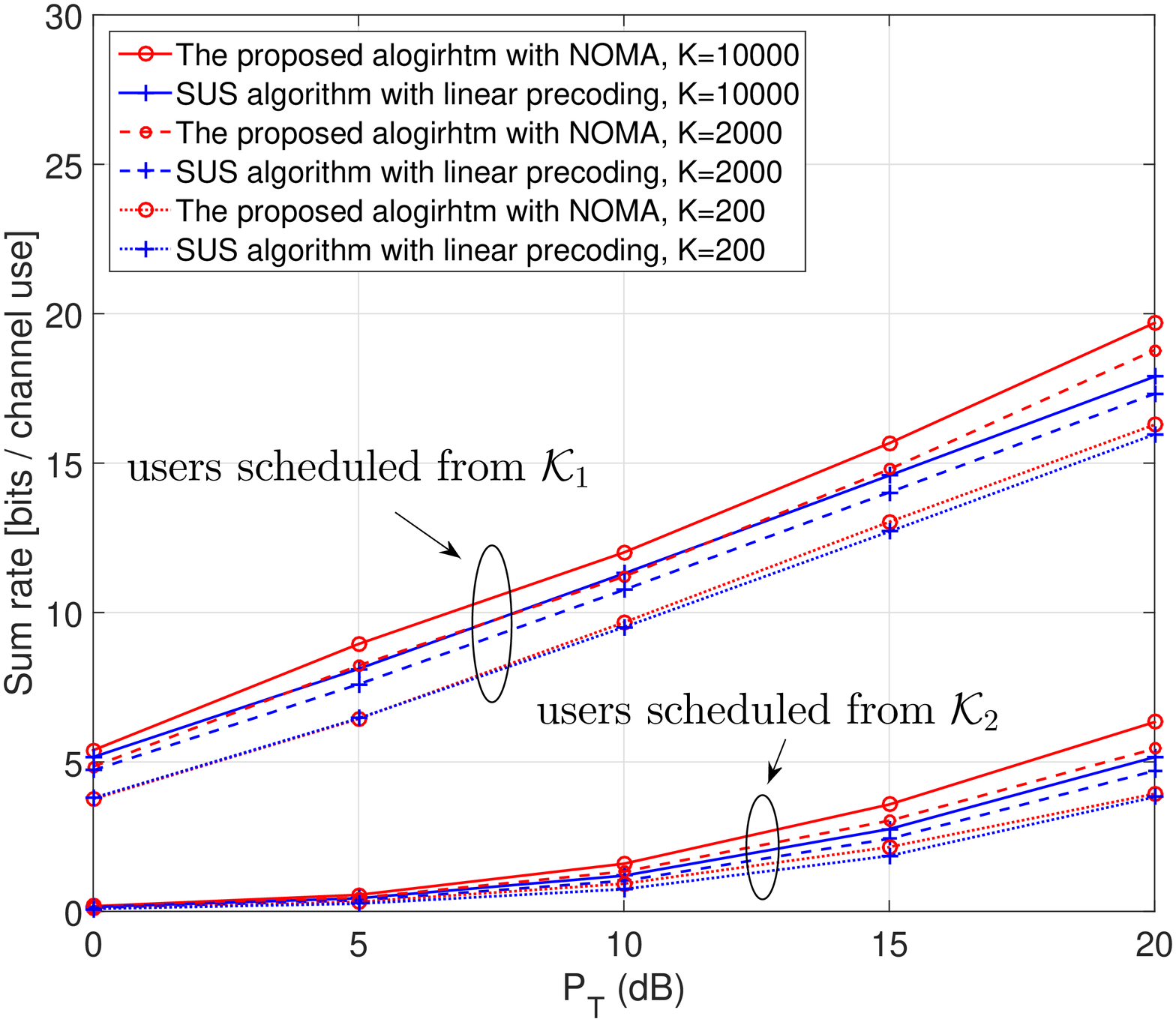} } }
\vspace{1.5em} \caption{Sum rate ($N_t = 4$) : (a) total sum rate and
(b) separate sum rates from $\Kc_1$ and from $\Kc_2$} \label{fig:sumrate_Nt4}
\end{figure*}

First, we evaluated the gain of the proposed method over conventional SUS-based MU-MISO scheduling  \cite{Yoo&Goldsmith}
and the results are shown in Figs. \ref{fig:sumrate_Nt2} and \ref{fig:sumrate_Nt4}.
The simulation setup for  Figs. \ref{fig:sumrate_Nt2} and \ref{fig:sumrate_Nt4} is as follows.  The AWGN variance was one for all users.
The numbers of transmit antennas were two and four for Fig.  \ref{fig:sumrate_Nt2} and Fig. \ref{fig:sumrate_Nt4}, respectively.
Each element of each channel vector in the strong user set $\Kc_1$ with $K/2$ users was randomly and independently generated from $\Cc\Nc(0, \sigma_{h,1}^2)$ with $\sigma_{h,1}^2=1$ and each element of each channel vector in the weak user set $\Kc_2$ with $K/2$ users was randomly and independently generated from $\Cc\Nc(0, \sigma_{h,2}^2)$ with $\sigma_{h,2}^2=0.01$. (Hence, we have $\lambda_1/\lambda_2=100=20$ dB.)
For the conventional SUS-based MU-MISO scheduling we considered two scheduling intervals. At the first interval, user scheduling out of $\Kc_1$ was performed by running the SUS algorithm for MU-MISO with $N_t$ transmit antennas and the scheduled users were served by ZF downlink beamforming\cite{Yoo&Goldsmith}. At the second interval,  user scheduling out of $\Kc_2$ was performed by running the SUS algorithm for MU-MISO with $N_t$ transmit antennas and the scheduled users were served by ZF downlink beamforming. The average sum rates for $
\Kc_1$ and $\Kc_2$ were obtained by averaging the rates of 1000 independent channel realizations. For the overall sum rate of the conventional SUS method, the two rates of $\Kc_1$ and $\Kc_2$ were averaged. For the proposed NOMA method, the same 1000 channel realizations used for the conventional method were used but the proposed NOMA scheduling was performed over the overall user set $\Kc_1
\cup \Kc_2$ in a single scheduling interval.  For the SUS algorithm applied separately to $\Kc_1$ and $\Kc_2$ and to Step 1 of the proposed method, the semi-orthogonality parameter $\delta$ should be chosen \cite{Yoo&Goldsmith} and we used optimal $\delta$ for each $K/2$ provided from Fig. 2 of \cite{Yoo&Goldsmith}. In addition, for the proposed method,  $\Gamma$ should be chosen and we set $\Gamma$ appropriately to balance the rates from the two groups $\Kc_1$ and $\Kc_2$.
It is seen in Figs. \ref{fig:sumrate_Nt2} and \ref{fig:sumrate_Nt4} that
 the proposed MU-MISO NOMA method outperforms the conventional MU-MISO downlink based on the SUS user scheduling. Note that the sum rates for both groups $\Kc_1$ and $\Kc_2$ are better than the conventional scheme.  It is also seen that the performance improvement by NOMA reduces as $N_t$ increases from two to four, but there exists non-trivial gain for NOMA for large $K$.

 \begin{figure}[!t]
\centering
\scalefig{0.5}\epsfbox{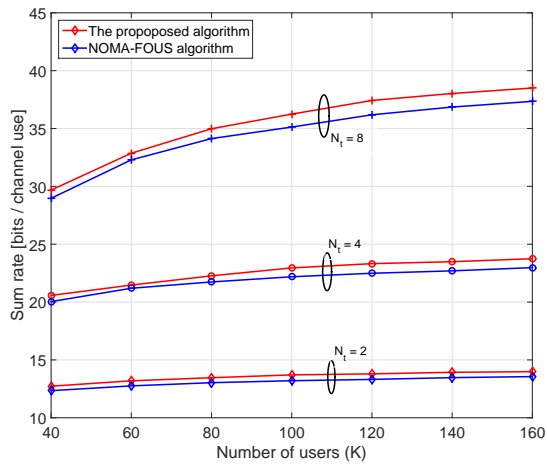}
\caption{Sum rate ($N_t = 2, 4, 8$, $K = 2000$ and $P_T = 15$dB)}
\label{fig:performance_comparison}
\end{figure}

Next, we compared the proposed algorithm with an existing
 algorithm proposed for MU-MISO NOMA downlink. For the comparison baseline we considered the NOMA-FOUS algorithm in \cite{Sayed17:WD} of which superiority over other methods \cite{Kim13:MILCOM,Liu15:ICCW} is shown in \cite{Sayed17:WD}.
Since the simulation setting is different from that in \cite{Sayed17:WD},
we slightly modified the NOMA-FOUS algorithm so that the strong user is selected from $\Kc_1$ and the weak user is selected from $\Kc_2$, although the original NOMA-FOUS algorithm considers only one set of users.
For comparison, we set $\sigma_{h,1}^2=1$ and $\sigma_{h,2}^2 = 0.04$, and
set $\Gamma$ to make the sum rate of the weak users of the proposed algorithm larger than the sum rate of the weak users
of the NOMA-FOUS algorithm. Fig. \ref{fig:performance_comparison} shows the
sum rate performance of the two algorithms.
It is seen that
the proposed algorithm outperforms the NOMA-FOUS algorithm.

 \begin{figure}[!t]
\centering
\scalefig{0.5}\epsfbox{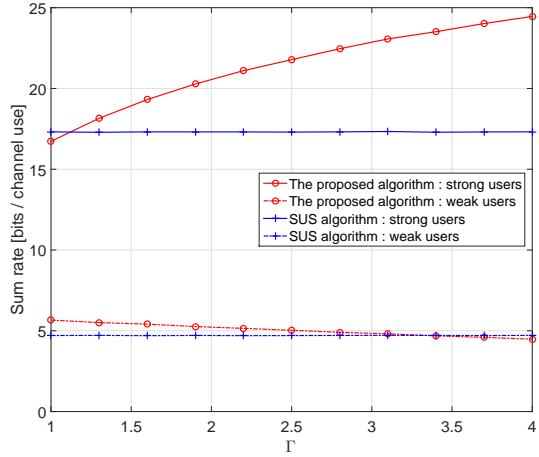}
\caption{Sum rate  versus $\Gamma$ ($N_t = 4$, $K = 2000$ and $P_T = 20$ dB)}
\label{fig:performance_Gamma}
\end{figure}

 The key feature of our Pareto-optimality-based design is that we have control over the rate operating point. Hence, we finally investigated the rate balancing property between the two groups $\Kc_1$ and $\Kc_2$ by controlling the strong-user-target-SINR parameter $\Gamma$ defined in \eqref{eq:channelSNRandGamma} (larger $\Gamma$ means larger rates for strong users), and the result is shown in Fig.  \ref{fig:performance_Gamma}. The simulation parameters are the same as those for Fig. \ref{fig:sumrate_Nt4}. For reference, the rates of $\Kc_1$ and $\Kc_2$ separately obtained by the conventional MU-MISO SUS algorithm are shown. It is seen that by abandoning the improvement for weak users but maintaining the weak-user performance at the level of the conventional SUS method, significant rate gain can be attained for strong users.

\section{Conclusion}
\label{sec:conclusion}

In this paper, we have considered the problem of transmit beam design and user scheduling  for MU-MISO NOMA downlink and proposed
 an effective  beam design and user scheduling method based on Pareto-optimality by exploiting both the spatial and power domains available in MU-MISO NOMA downlink. The proposed method with the ability of rate control between strong and weak users provides great flexibility to NOMA network operation.

\section*{Appendix}

{\em Proof of Proposition \ref{pro:PcregionDetermine}:}  ~ The set
$\Pc_i$ to which $p_1^{opt}$ belongs  is dependent on the
relationship among $a(p_1), b(p_1)$ and $d(p_1):=b(p_1) +
c^2(p_1)/b(p_1)$, given in terms of $\Gamma$, $\theta$ and
$\lambda_i$ by
    \begin{align}
        a(p_1) &=   \sqrt{P - p_1} \sqrt{\frac{ \|\hbf_1\|^2}{\sigma_1^2 (1+ \gamma_1^*)}} = \sqrt{(P-p_1) \frac{\lambda_1}{1 + \Gamma \lambda_1}}  \label{eq:a_p1append}\\
        b(p_1) &=   \sqrt{P - p_1} \sqrt{\frac{\|\hbf_2\|^2 \theta}{\|\hbf_2\|^2 p_1 [\alpha_1^{*}(p_1)]^2 + \sigma_2^2}} \nonumber\\
        &= \sqrt{(P-p_1) \frac{\lambda_2 \theta}{\lambda_2 p_1 [\alpha_1^{*}(p_1)]^2+ 1} } \label{eq:b_p1append} \\
        d(p_1) &= \sqrt{P - p_1} \sqrt{\frac{ \|\hbf_2\|^2 /\theta}{{\|\hbf_2\|^2 p_1 [\alpha_1^{2}(p_1)]^2 + \sigma_2^2}}} \nonumber \\
        &=  \sqrt{(P-p_1) \frac{\lambda_2/\theta}{\lambda_2 p_1  [\alpha_1^{*}(p_1)]^2 + 1}}. \label{eq:c_p1append}
    \end{align}
The three sets $\Pc_1$, $\Pc_2$ and $\Pc_3$ can be rewritten by
squaring $a(p_1)$, $b(p_1)$ and $d(p_1)$ and dropping the common
factor $(P-p_1)$ as $\mathcal{P}_1 = \{ p_1| \bar{a} \le
\bar{b}(p_1) \}$, $\mathcal{P}_2 = \{ p_1| \bar{b}(p_1) < \bar{a}
\leq \bar{d}(p_1) \}$, and $\mathcal{P}_3 = \{ p_1| \bar{a} >
\bar{d}(p_1) \}$, where
\begin{align} \label{eq:barabcinproof}
&\bar{a} =  \frac{\lambda_1}{1 + \Gamma \lambda_1},~ \bar{b}(p_1) =
\frac{\lambda_2 \theta}{\lambda_2 p_1 [\alpha_1^{*}(p_1)]^2+ 1}~, ~\mbox{and}~ \\
   & \bar{d}(p_1) = \frac{\lambda_2/\theta}{\lambda_2 p_1
[\alpha_1^{*}(p_1)]^2 + 1}.\label{eq:barabcinproof2}
\end{align}

First, we show  $p_1^{opt} \notin \mathcal{P}_1$. Let $p_{1,min}$
denote the minimum $p_1$ to achieve $\gamma_1^*$ with
$\wbf_1=\hbf_1/||\hbf_1||$. Then, $p_{1,min} =
\gamma_1^*/\lambda_1 = \Gamma$. Hence, the second condition in
\eqref{eq:x_1_opt_with_p1}, i.e., $\Gamma > p_{1,min}(1-\theta)$
or $\Gamma = p_{1,min}(1-\theta)$ is satisfied since  $0 \le
\theta \le 1$, and $\alpha_1^*(p_{1,min})=\sqrt{\theta}$ from
\eqref{eq:x_1_opt_with_p1}. Hence, we have
\begin{align}
       \bar{a} &=  \frac{\lambda_1}{1+ \Gamma\lambda_1} =  \frac{1}{\frac{1}{\lambda_1}+
       \Gamma}  \\
         \bar{b}(p_{1,min}) &=  \frac{\lambda_2 \theta}{\lambda_2 \Gamma \theta  + 1} =  \frac{1}{\frac{1}{\theta \lambda_2}+ \Gamma}.
\end{align}
By the NOMA condition $\lambda_1 > \lambda_2$, we have
$\frac{1}{\lambda_1} < \frac{1}{\theta \lambda_2}$ since $0 \le
\theta \le 1$ and thus $\bar{a} > \bar{b}(p_{1,min})$. In case of
$\Gamma = p_{1,min}(1-\theta)$, we have $\theta=0$ and thus
$\bar{b}(p_{1,min})=0$ and $\bar{a} > \bar{b}(p_{1,min})$. Hence,
$ p_{1,min} \notin \mathcal{P}_1$. Note that $\bar{a}$ is constant
over $p_1$. It can be shown from
  \eqref{eq:x_1_opt_with_p1} that the term $p_1[\alpha_1^*(p_1)]^2$ in the
denominator of $\bar{b}(p_1)$ in \eqref{eq:barabcinproof} is
monotone decreasing with respect to $p_1$, and hence
$\bar{b}(p_1)$ is monotone increasing with respect to $p_1$. If
$\bar{a} > \bar{b}(p_1)$ for all $p_1$, $\Pc_1$ is empty.
Otherwise, there exists
 $p_1$, denoted as  $p_{1,a}$, such that $\bar{a}
= \bar{b}(p_1)$, as $p_1$ increases, given by
{\small{
\begin{eqnarray}
        p_{1,a} &=& \{ p_1 | \bar{a} = \bar{b}(p_1)\} \nonumber \\
        &=& \{ p_1 | \frac{\lambda_1}{1 + \Gamma \lambda_1}  = \frac{\lambda_2 \theta}{\lambda_2 p_1  [\alpha_1^{*}(p_1)]^2 + 1}\} \nonumber \\
                &=& \Gamma + \frac{1}{1 - \theta}\left(\sqrt{\theta \Gamma} - \sqrt{\theta \Gamma + \lambda_1^{-1} \theta - \lambda_2^{-1}}\right)^2. \label{eq:p_1a}
\end{eqnarray}
}}
At $p_1=p_{1,a}$, we have $\gamma_2^{*(2)}=\gamma_2^{*(1)}$ from
\eqref{eq:gamma2_p1} since $\alpha_2^*(p_1)=1$ at the boundary of
$\Pc_1$ ($p_1 \ge p_{1,a}$ side) and $\Pc_2$ ($p_1 < p_{1,a}$
side), i.e., $\bar{a}=\bar{b}(p_1)$. Furthermore, it can be shown
that
    $ \left.\frac{\partial \gamma_2^{*(2)}(p_1)}{\partial p_1}\right|_{p_1 \rightarrow p_{1,a}^-} =
    \frac{-1}{c_1}$,
     where $c_1$ is a non-negative constant with respect to $p_1$.
    Hence, there exists $p_1 \in \mathcal{P}_2$ such that $\gamma_2^{*(2)}(p_1) > \gamma_2^{*(2)}(p_{1,a})= \gamma_2^{*(1)}(p_{1,a})$.
Since  $\gamma_2^{*(1)}$ is a monotone decreasing function of
$p_1$ as seen in \eqref{eq:gamma2_p1}, optimal $\gamma_2^*$ does
not occur in $\Pc_1$, i.e.,  $p_1^{opt} \notin \mathcal{P}_1$.

Next, we check the condition that $\Pc_3$ is empty.  Since the
term $p_1[\alpha_1^*(p_1)]^2$ in the denominator of $\bar{d}(p_1)$
in \eqref{eq:barabcinproof2} is monotone decreasing with respect to
$p_1$, and thus $\bar{d}(p_1)$ is monotone increasing with respect
to $p_1$. Therefore, if $\bar{a} < \bar{d}(p_{1,min})$, then
$\Pc_3$ is empty.   Since $\alpha_1^*(p_{1,min})=\sqrt{\theta}$
from \eqref{eq:x_1_opt_with_p1} and $p_{1,min}=\Gamma$, the
condition is rewritten from \eqref{eq:barabcinproof} and \eqref{eq:barabcinproof2} as
    \begin{eqnarray}
         \bar{a} < \bar{d} (p_{1,min}) &\Leftrightarrow&  \frac{\lambda_1}{1+ \Gamma\lambda_1} <
           \frac{\lambda_2/\theta}{\lambda_2\Gamma \theta +1} \\
           &\Leftrightarrow& \theta \Gamma < \frac{1}{\theta} \left( \frac{1}{\lambda_1} +\Gamma \right) - \frac{1}{\lambda_2} =:
   \tau.
    \end{eqnarray}
    In this case, $\Pc_3 = \emptyset$ and  $p_1^{opt} \in \mathcal{P}_2$ since $p_1^{opt} \notin
    \Pc_1$.

Now assume $\theta \Gamma \ge \tau$. Then, $\Pc_3$ is not
    empty. Furthermore, we have a sufficient condition for
    $\forall~ p_1 \in \Pc_3$ as follows:
    \begin{eqnarray}
        \bar{d}(p_1) < \bar{a}, \quad \forall p_1 &\Leftarrow&  \lambda_2/\theta < \frac{\lambda_1}{1 + \Gamma \lambda_1}  \\
                                                    &\Leftrightarrow&  \frac{1}{\theta} \left( \frac{1}{\alpha_1} +\Gamma \right) - \frac{1}{\lambda_2}  < 0 \\
                                                    &\Leftrightarrow& \tau <
                                                    0,
    \end{eqnarray}
    because $\lambda_2/\theta$ is an upper bound of $\bar{d}(p_1)$
    (see \eqref{eq:barabcinproof2}).  In this case, $p_{1}^{opt}
    \in \Pc_3$.

Finally, if $\theta \Gamma \geq \tau$ and $\tau \geq 0$, compute
$p_1$, denoted by $p_{1,b}$, such that $\bar{a} = \bar{d}(p_1)$,
given by
    \begin{eqnarray}
        p_{1,b} &=& \{ p_1 | \bar{a} = \bar{d}(p_1)\} \\
                &=& \{ p_1 | \frac{\lambda_1}{1 + \Gamma \lambda_1}  = \frac{\lambda_2/\theta}{\lambda_2 p_1  [\alpha_1^{*}(p_1)]^2 + 1}\} \\
                &=& \Gamma + \frac{1}{1 - \theta}\left(\sqrt{\theta \Gamma} -
                \sqrt{\tau}\right)^2.\label{eq:proofProp1p1blast}
    \end{eqnarray}
If
\begin{equation}  \label{eq:proofProp1Pp1b}
P< p_{1,b},
\end{equation}
 then $\bar{a} > \bar{d}(p_1)$ for $p_1 \le P$ since
$\bar{d}(p_1)$ is a monotone increasing function of $p_1$. Hence,
in this case, $\forall p_1 \in \Pc_3$ and $p_1^{opt} \in \Pc_3$.
On the other hand, if $p_{1,b} \le P$, we have both nonempty
$\Pc_2=\{p_1 \ge p_{1,b}\}$ and $\Pc_3=\{p_1 < p_{1,b}\}$. In this
case, we compute the derivatives  of $\gamma_2^{*(2)}$ and
$\gamma_2^{*(3)}$ at point $p_{1,b}$, which are given by

    \begin{align}
        \left.\frac{\partial \gamma_2^{*(2)}}{\partial p_1}\right|_{p_1 = p_{1,b}^+}
        = c_2 \left[\left( \lambda_2 \sqrt{\tau} \frac{1-\theta}{\sqrt{\theta \Gamma}-\sqrt{\tau}}\right)P \right. 
        &- \left.\left( \lambda_2 \sqrt{\tau}
        \frac{1-\theta}{\sqrt{\theta \Gamma}-\sqrt{\tau}} \cdot \Gamma + \lambda_2\sqrt{\tau} \cdot \sqrt{\theta
        \Gamma} + 1 \right)\right] \\
        \left.\frac{\partial \gamma_2^{*(3)}}{\partial p_1}\right|_{p_1 = p_{1,b}^-}
        = c_3 \left[ \left( \lambda_2 \sqrt{\tau} \frac{1-\theta}{\sqrt{\theta \Gamma}-\sqrt{\tau}}\right) P \right. 
        &- \left.\left( \lambda_2 \sqrt{\tau}
        \frac{1-\theta}{\sqrt{\theta \Gamma}-\sqrt{\tau}} \cdot \Gamma + \lambda_2\sqrt{\tau} \cdot \sqrt{\theta \Gamma} + 1
        \right)\right],
    \end{align}
    \hspace{-0.5em}where $c_2$ and $c_3$ are non-negative constants. The two derivatives have the same sign. If the two derivatives are positive,
    then $\gamma_2^*$ increases as $p_1$ crosses $p_{1,b}$ from
    the left to the right and hence $p_1^{opt}\in \Pc_2$.
    Otherwise, $\gamma_2^*$ increases as $p_1$ crosses $p_{1,b}$ from
    the right to the left and hence $p_1^{opt}\in \Pc_3$.
    Equivalently, we have
    \begin{align}
       p_1^{opt} \in \mathcal{P}_2 &~~\text{if}~~  P \geq \Gamma + \frac{1}{1- \theta} (\sqrt{\theta \Gamma}
          - \sqrt{\tau})(\sqrt{\theta \Gamma} + \frac{1}{\lambda_2 \sqrt{\tau}} )  \label{eq:proofProp1lasteq2}\\
       p_1^{opt} \in \mathcal{P}_3 &~~\text{if}~~  P < \Gamma + \frac{1}{1- \theta} (\sqrt{\theta \Gamma} -
           \sqrt{\tau})(\sqrt{\theta \Gamma} + \frac{1}{\lambda_2 \sqrt{\tau}}
          ).  \label{eq:proofProp1lasteq}
    \end{align}
Since  $p_{1,b}=\Gamma + \frac{1}{1 - \theta}\left(\sqrt{\theta
\Gamma} -                \sqrt{\tau}\right)^2 < \Gamma +
\frac{1}{1- \theta} (\sqrt{\theta \Gamma} -
           \sqrt{\tau})(\sqrt{\theta \Gamma} + \frac{1}{\lambda_2 \sqrt{\tau}}
          )$, the set $\{P< p_{1,b}\}$ mentioned in
          \eqref{eq:proofProp1Pp1b} is a subset of the set $\{P < \Gamma + \frac{1}{1- \theta} (\sqrt{\theta \Gamma} -
           \sqrt{\tau})(\sqrt{\theta \Gamma} + \frac{1}{\lambda_2 \sqrt{\tau}}
          )\}$. Thus, the case of $P< p_{1,b}$ is
            covered by
          \eqref{eq:proofProp1lasteq}.  The only two cases for
          $p_1^{opt} \in \Pc_2$ are [$\theta\Gamma < \tau$] or
          [$\theta \Gamma \ge \tau \ge 0$ and $P \geq \Gamma + \frac{1}{1- \theta} (\sqrt{\theta \Gamma}
          - \sqrt{\tau})(\sqrt{\theta \Gamma} + \frac{1}{\lambda_2 \sqrt{\tau}} )
          ]$.  Hence,        the claim follows.
    \hfill{$\square$}

\vspace{1em}
 To prove Proposition
\ref{pro:proposition1Angle}, we introduce the following lemma.

   \begin{lemma} \label{lem:lemma1} Define
     \begin{eqnarray}
        \mathcal{N}_1 &:=& \{ \theta \; | \; a \leq b(\theta) \}  \label{eq:lemNc1} \\
        \mathcal{N}_2 &:=& \{ \theta \; | \; b(\theta) < a \leq b(\theta)+ c^2(\theta)/b(\theta) \} \\
        \mathcal{N}_3 &:=& \{ \theta \; | \; a > b(\theta) + c^2(\theta)/b(\theta) \}, \label{eq:lemNc3}
    \end{eqnarray}
    where $a$, $b(\theta)$ and $c(\theta)$ are defined just below \eqref{eq:PO_objective_6}.
    Then, for  given $\lambda_1$, $\lambda_2$ and $\Gamma$,  every $\theta$ in $\mathcal{N}_1$ achieves the same optimal $\gamma_2^*(\theta)$ in \eqref{eq:gamma_2_fixed_power_eta}, if $\Nc_1$ is not empty. \label{lemma1}
    \end{lemma}

    {\em Proof:} ~~~       Let us assume that $\Nc_1$ is not empty.  For every $\theta \in \Nc_1$, $\gamma_2^*(\theta)= \gamma_2^{*(1)}== \frac{\lambda_1}{1 + \lambda_1 \Gamma}$.
        From the fact that
           $\gamma_2^{*(1)} = \frac{\lambda_1}{1 + \lambda_1 \Gamma} ~~\mbox{and}~~                \gamma_2^{*(2)}(\theta) = \frac{\lambda_1}{1 + \lambda_1 \Gamma} [\alpha_2^{*}(\theta)]^2$,
                    and $0 \leq \alpha_2^{*}(\theta) < 1$ for $\theta \in \Nc_2$, it is obvious that $ \gamma_2^{*(2)}(\theta) < \gamma_2^{*(1)}$ for all $\theta \in \Nc_2$. Hence, $\gamma_2^*(\theta)$ for $\theta \in \Nc_2$ is less than $\gamma_2^*(\theta)$ for $\theta \in \Nc_1$.
                    Furthermore, for any $\theta \in \mathcal{N}_3$, we have
                        $\gamma_2^{*(3)}(\theta) \stackrel{(a)}{=} \frac{ \lambda_2 }{ \lambda_2 [\alpha_1^{*}(\theta)]^2 + 1}                                   \stackrel{(b)}{\leq} \frac{ \lambda_2/\theta }{ \lambda_2 [\alpha_1^{*}(\theta)]^2 + 1} \stackrel{(c)}{=} [b(\theta) + c^2(\theta)/b(\theta)]^2  \stackrel{(d)}{<} a^2 \stackrel{(e)}{=} \gamma_2^{*(1)}$.
                        Here, step (a) is by \eqref{eq:gamma_2_fixed_power}, step (b) holds because $\theta \in [0,1]$, step (c) is by direction computation based on $b(\theta)$ and $c(\theta)$,  step (d) holds because $\theta \in \Nc_3$,  and step (e) is by \eqref{eq:gamma_2_fixed_power}.
            Consequently, we have the claim.   \hfill{$\square$}


\vspace{1em}

{\em Proof of Proposition \ref{pro:proposition1Angle}:} ~~ The necessary and sufficient condition for $\Nc_1$ defined in \eqref{eq:lemNc1} being non-empty is given by
    \begin{equation}  \label{eq:lem1Nc1condition}
        a^2 \leq \max_{0 \leq \theta \leq 1} b^2(\theta),
    \end{equation}
    where
    {\small{
    \begin{equation} \label{eq:appendProp2abc}
    a^2=\frac{\lambda_1}{1+\Gamma \lambda_1}, ~~b^2(\theta)= \frac{\lambda_2\theta}{\lambda_2[\alpha_1^*(\theta)]^2+1}~~c^2(\theta)= \frac{\lambda_2(1-\theta)}{\lambda_2[\alpha_1^*(\theta)]^2+1}.
    \end{equation}
    }}
    Since $b(\theta)$ is maximized at
    $\theta = \frac{[\lambda_2(1-\Gamma) +1]^2}{\lambda_2^2 \Gamma(1-\Gamma) + [\lambda_2(1-\Gamma) +1]^2}$
    and the corresponding maximum value is   $\max_{0 \leq \theta \leq 1}$ $b^2(\theta)  = \lambda_2 (1 + \frac{\lambda_2 \Gamma }{\lambda_2 + 1})$,
 the condition \eqref{eq:lem1Nc1condition} becomes

    \begin{eqnarray}
       && \frac{\lambda_1}{1 + \lambda_1 \Gamma} \leq \lambda_2 (1 + \frac{\lambda_2 \Gamma }{\lambda_2 + 1}) \nonumber \\
       &\Leftrightarrow& \Gamma_1:=\frac{1}{2} \left(  1 +  \lambda_{2}^{-1}-\lambda_{1}^{-1} \right)- \frac{1}{2} \sqrt{( 1 + \lambda_{1}^{-1} + \lambda_{2}^{-1})^2 - 4\lambda_{2}^{-1}(1 + \lambda_{2}^{-1})} ~\leq \Gamma \nonumber \\
        && \leq  \frac{1}{2} \left(  1 +  \lambda_{2}^{-1}-\lambda_{1}^{-1} \right) + \frac{1}{2} \sqrt{( 1 + \lambda_{1}^{-1}
        + \lambda_{2}^{-1})^2 - 4\lambda_{2}^{-1}(1 +
       \lambda_{2}^{-1})} =:\Gamma_2.\nonumber
    \end{eqnarray}

 In the case of non-empty $\Nc_1$,  by substituting $\alpha_1^*(\theta)$ in  \eqref{eq:x_1_opt} or \eqref{eq:appendProp2proofalpha1star} into $b^2(\theta)$, $b^2(\theta)$ is
 given by
 \begin{equation}
    b^2(\theta) = \left\{\begin{array}{cc}
                          \lambda_2\theta & ~~ \mbox{if} ~~  \theta \le 1-\Gamma \\
                          \frac{\lambda_2\theta}{\lambda_2[\sqrt{\theta\Gamma} - \sqrt{(1-\theta)(1-\Gamma)}]^2+1} & ~~ \mbox{if} ~~ \theta > 1- \Gamma
                        \end{array}
     \right.
 \end{equation}
When $\theta \le 1-\Gamma$, $b^2(\theta)$ is linear and it can be shown that $b^2(\theta)$ is quasi-concave function when $\theta > 1-\Gamma$
 \footnote{ When $\theta > 1-\Gamma$, $b^2(\theta)$ can be written as $f^2(\theta)/g(\theta)$, where $f(\theta) = \sqrt{\lambda_2 \theta}$ and $g(\theta) = \lambda_2[\sqrt{\theta\Gamma} - \sqrt{(1-\theta)(1-\Gamma)}]^2+1$. Since $f(\theta)$ is the concave function and $g(\theta)$ is the convex function (it can be proved easily by taking secondary derivative), we can conclude that
  $f^2(\theta)/g(\theta)$ is quasi concave\cite{Bector:68OR}.}.
  If $a^2 > \lambda_2 (1-\Gamma)$, there doesn't exists $\theta$ satisfying $a^2  \leq \lambda_2 \theta $ (for $\theta \leq 1 -\Gamma$) and hence the set $\Nc_1=\{ \theta \; | \; a^2 \leq b^2(\theta) \}$ is given by

     \begin{eqnarray}
          && \biggr\{\theta~|~\frac{z_1 z_2 + 2\Gamma(1-\Gamma) - \sqrt{4\Gamma(1-\Gamma) [\Gamma(1-\Gamma) + z_1z_2 -z_2^2]}}{z_1^2 + 4\Gamma(1-\Gamma)}  \nonumber  \\
           && ~~~~\leq \theta   \leq  \frac{z_1 z_2 + 2\Gamma(1-\Gamma) + \sqrt{4\Gamma(1-\Gamma) [\Gamma(1-\Gamma) + z_1z_2 -z_2^2]}}{z_1^2 +
           4\Gamma(1-\Gamma)}\biggr\} \label{eq:appendProp2pfsecondset}
    \end{eqnarray}

    Otherwise, the minimum of $\theta$ satisfying $a^2 \leq b^2(\theta)$ is the point such that $a^2 = \lambda_2 \theta$ and $\Nc_1$ becomes
  
    \begin{eqnarray}
          &&\left\{ \theta ~| ~\frac{\lambda_1}{\lambda_2} \frac{1}{1 + \lambda_1  \Gamma }  \leq \theta \leq \frac{z_1 z_2 + 2\Gamma(1-\Gamma) + \sqrt{4\Gamma(1-\Gamma) [\Gamma(1-\Gamma) + z_1z_2 -z_2^2]}}{z_1^2 +
           4\Gamma(1-\Gamma)}\right\}, \nonumber
    \end{eqnarray}

    where $z_1=\lambda_{1}^{-1} + 1-\Gamma$ and $z_2 =
\lambda_{2}^{-1} + 1-\Gamma$. (\eqref{eq:appendProp2pfsecondset}
is obtained by solving $ \frac{\lambda_1}{1+\Gamma \lambda_1} \leq \frac{ \lambda_2\theta}{\lambda_2[\sqrt{\theta\Gamma} - \sqrt{(1-\theta)(1-\Gamma)}]^2+1}$
reducing to a quadratic inequality).  In the case of non-empty $\Nc_1$, by Lemma
\ref{lem:lemma1}, $\Nc_1$ is optimal and we obtain \eqref{eq:opt_eta}.

    Next, consider the case that $\Nc_1$ is empty.  At $\theta= 1$, we have
    $        a(1) = \sqrt{\frac{1}{\Gamma + \frac{1}{\lambda_1}}}    >  \sqrt{\frac{1}{\Gamma + \frac{1}{\lambda_2}}}
             = b(1) + c^2(1)/b(1)$ by the NOMA assumption $\lambda_1 > \lambda_2$ and hence
    $\theta= 1 \in \mathcal{N}_3$ by the definition of $\Nc_3$ in \eqref{eq:lemNc3}.
    We also have
    \begin{equation}
        \lim_{\theta \rightarrow 0} b(\theta) + c^2(\theta)/b(\theta)  = \lim_{\theta \rightarrow 0} \sqrt{\frac{\lambda_2\theta^{-1}}{\lambda_2[\alpha_1^*(\theta)]^2+1}}=\infty,
    \end{equation}
which can easily be seen from  $\alpha_1^*(0)=\sqrt{1-\Gamma}$.
Thus, $\theta=0 \in \Nc_2$. Furthermore, $b(\theta) +
c^2(\theta)/b(\theta)=\sqrt{\frac{\lambda_2\theta^{-1}}{\lambda_2[\alpha_1^*(\theta)]^2+1}}$
is a  monotone decreasing function of $\theta$ since
    {\small{
   \begin{equation}  \label{eq:appendProp2proofalpha1star}
        \alpha_1^*(\theta) = \left\{ \begin{array}{cl}
                          0 & \quad   \text{if} \quad \theta \leq \theta_I:=1-\Gamma   \\
                          \sqrt{\theta \Gamma} - \sqrt{(1-\theta) (1 - \Gamma )} & \quad  \text{if} \quad  \theta > \theta_I
                        \end{array}
        \right.
    \end{equation}}}
    is a monotone increasing function of  $\theta$.
    Hence,
    there exists $\theta_a$ such that $a(\theta_a) = b(\theta_a) +
    c^2(\theta_a)/b(\theta_a)$
    to yield $\Nc_2=\{\theta|\theta \le \theta_a \}$ and $\Nc_3=\{\theta|\theta > \theta_a\}$.

Now recall that
           \begin{eqnarray}
          && \gamma_2^{*(1)} = \frac{\lambda_1}{1 + \lambda_1
           \Gamma},~~
           \gamma_2^{*(2)}(\theta) = \frac{\lambda_1}{1 + \lambda_1 \Gamma}
           [\alpha_2^{*}(\theta)]^2,  ~\mbox{and}~  \gamma_2^{*(3)}(\theta) = \frac{ \lambda_2 }{ \lambda_2 [\alpha_1^{*}(\theta)]^2 +
                        1}.\label{eq:proofProp2gamma2theta}
           \end{eqnarray}
    If  $\theta_a \leq \theta_I$, then the optimal
    $\theta$ set for maximizing $\gamma_2^*$ is given by $\{\theta|\theta_a \leq \theta \leq
    \theta_I\}$. This is because $\gamma_2^{*(2)}(\theta_a)=\gamma_2^{*(3)}(\theta_a)$,
    because
     $\gamma_2^{*(3)}(\theta)$ is monotone decreasing with respect to
     $\theta$ as seen in \eqref{eq:proofProp2gamma2theta}
     since $\alpha_1^*(\theta)$ is a monotone increasing function of
    $\theta$, and because
$\gamma_2^{*(2)}(\theta)$ is monotone increasing with respect to
$\theta$ for $\theta \le \theta_a$ since  $\alpha_2^*(\theta)$ is
a monotone increasing function of
    $\theta$ for $\theta \le \min\{\theta_a,\theta_I\}=\theta_a$ (this can be shown by substituting $\alpha_1^*=0$ for  $\theta \le \theta_I$
    into $\alpha_2^*(\theta)$ and taking derivative of  $\alpha_2^*(\theta)$ with respect to $\theta$ and showing the derivative is positive for
     $\theta \le \theta_a$).
    Hence, in this case the optimal $\gamma_2^*$ occurs at $\theta_a$ but for all $\theta$ in $\{\theta|\theta_a \leq \theta \leq
    \theta_I\}$, $\alpha_1^*(\theta)=0$ and the corresponding
    optimal $\gamma_2^*=\gamma_2^{*(3)}=\lambda_2$ from
    \eqref{eq:proofProp2gamma2theta}. In this case, from the
    assumption $\theta_a \leq \theta_I$, $\theta_a$ is computed based on \eqref{eq:appendProp2abc} with $\alpha_1^*(\theta)=0$ as
   \begin{eqnarray}
        \theta_a &=& \theta ~~\mbox{s.t.}~~ a = b(\theta) + c^2(\theta)/b(\theta) \\
                &=& \theta ~~\mbox{s.t.}~~   \frac{\lambda_1}{1 + \lambda_1 \Gamma} = \frac{\lambda_2}{\theta} \\
                &=& \frac{\lambda_2}{\lambda_1} (1 + \lambda_1 \Gamma)
    \end{eqnarray}
     and the condition  $\theta_a \leq \theta_I$ reduces to
     \begin{equation}
        \frac{\lambda_2}{\lambda_1} (1 + \lambda_1 \Gamma) \leq 1 - \Gamma
        ~\Longleftrightarrow~ \Gamma \leq \frac{ {\lambda_2}^{-1} - {\lambda_1}^{-1}}{1 + {\lambda_2}^{-1}}.
     \end{equation}
     On the other hand, if $\theta_a > \theta_I$, i.e., $\Gamma > \frac{ {\lambda_2}^{-1} - {\lambda_1}^{-1}}{1 +
     {\lambda_2}^{-1}}$, then  optimal $\theta$ exists  between
$\theta_I$ and $\theta_a$ because  $\gamma_2^{*(2)}$ is an
increasing function for $\theta < \min\{\theta_a,\theta_I\}
=\theta_I$  and $\gamma_3^{*(3)}$ is a decreasing function for
$\theta > \theta_a$. Since optimal $\theta$ lies in $\Nc_2$ in
this case, it is obtained by solving
     \begin{equation}
        \frac{\partial \gamma_2^{*(2)}(\theta)}{\partial \theta} = 0.
     \end{equation}
Therefore, the claim follows. \hfill{$\square$}

\vspace{1em}
{\em Proof of Corollary \ref{pro:corollaryAngle}:} ~~    With the assumption of  $\lambda_1 = \lambda_2$, we have $\Gamma_1 = 0$ and $\Gamma_2 = 1$ from \eqref{eq:Gamma1and2} and hence
      the condition $\Gamma \in [\Gamma_1,\Gamma_2]$ reduces to  $\Gamma \in [0, 1]$ which is always valid
      for $p_1=p_2=1$ (see the definition of $\Gamma$ in \eqref{eq:channelSNRandGamma}). Furthermore, with the assumption, we have  $z_1 = z_2$ and $\sqrt{4\Gamma(1-\Gamma) [\Gamma(1-\Gamma) + z_1z_2 -z_2^2]}$ in \eqref{eq:opt_eta} is given by $2\Gamma(1-\Gamma)$. From \eqref{eq:opt_eta},  the optimal $\theta$ set  is given by
      $\{\theta ~|~ \theta_0 \le \theta \le 1\}$,  where
               \begin{equation}
        \theta_0 = \left\{~ \begin{array}{ll}
                              \frac{1}{1 + \Gamma \lambda_{1}}  ~~& \text{ if }~~
                               \frac{1}{1 + \Gamma \lambda_{1}} \leq 1 -\Gamma \\
                              \frac{z_1^2}{z_1^2 + 4\Gamma(1-\Gamma)} ~~& \text{ if }~~   \frac{1}{1 + \Gamma \lambda_{1}}  > 1 -\Gamma
                            \end{array}
          \right.
     \end{equation}
   \hfill{$\square$}

{\em Proof of Proposition \ref{prop:l1ggl2}:} ~~  As $\lambda_2 \rightarrow 0$.  the threshold $\tau$  in Proposition \ref{pro:PcregionDetermine} converges to $
   -\infty$ and thus neither  of the two conditions for $p_1^{opt} \in \Pc_2$ in Proposition \ref{pro:PcregionDetermine}    is satisfied. Hence, by Proposition  \ref{pro:PcregionDetermine},  $p_1^{opt} \in \Pc_3$. Since $p_1^{opt} \in \Pc_3$, $p_1^{opt}$ can
    be obtained in closed form by maximizing $\gamma_2^{*(3)}(p_1)$ in \eqref{eq:gamma2_p1} and is given by
    \begin{equation}
        \bar{p}_1^{opt} = -P + 2\Gamma +  \frac{\psi_1^2
- \psi_1\sqrt{ \psi_2^2 + 2\lambda_2^{-1}\psi_1 + \lambda_2^{-2} }}{2\theta(1-\theta)\Gamma }
    \end{equation}
    where $\psi_1 := \theta\Gamma  + (1-\theta)(P-\Gamma ) + \lambda_2^{-1}$ and $\psi_2 := \theta\Gamma  - (1-\theta)(P-\Gamma )$.
    Using L'Hospital's rule, we can show that
       $\lim_{\lambda_2 \rightarrow 0 }\bar{p}_1^{opt} = \Gamma (=p_{1,min})$.
     With $p_1 = \Gamma$, we have $\alpha_1^* = \sqrt{\theta}$ from \eqref{eq:x_1_opt_with_p1} and consequently $\beta_1^*=\sqrt{1-\theta}$ from  the the constraint eq. in \eqref{eq:PO_objective_4},
     and $\alpha_2^* = \sqrt{\theta}$ from \eqref{eq:x_2_optimal_with_p1}, and by substituting these values into $\gamma_2^{*(3)}(p_1)$ in \eqref{eq:gamma2_p1}  and \eqref{eq:parameterization1} and \eqref{eq:parameterization2}, we have  $\gamma_2^* = \frac{P-\Gamma}{\Gamma} \frac{1}{\theta + \lambda_2^{-1} \Gamma^{-1}}$,
    $\sqrt{p_1}\wbf_1 = \sqrt{\Gamma} \frac{\hbf_1}{\|\hbf_1\|}$  and
        $\sqrt{p_2}\wbf_2 = \sqrt{P-\Gamma} \frac{\hbf_2}{\|\hbf_2\|}$.
    \hfill{$\square$}

\bibliographystyle{ieeetr}



\end{document}